\documentclass[%
reprint,
 amsmath,amssymb,
 aps,
 twocolumn,
longbibliography,
]{revtex4-1}

\usepackage[T1]{fontenc}
\usepackage[dvipdfmx]{graphicx}
\usepackage{dcolumn}
\usepackage{bm}
\usepackage[driverfallback=dvipdfmx,
colorlinks=true,
linkcolor=blue,
filecolor=blue,
citecolor=blue,
urlcolor=blue
]{hyperref}

\usepackage{comment}

\usepackage{color}
\newcommand{\blue}[1]{\textcolor{blue}{#1}}
\usepackage[caption=false]{subfig}

\begin{document}

\title{Continuous-variable quantum approximate optimization on a programmable photonic quantum processor}

\author{Yutaro Enomoto}
 \email{yenomoto.ap.t@gmail.com}
 \affiliation{%
 Department of Applied Physics, School of Engineering, The University of Tokyo,\\
 7-3-1 Hongo, Bunkyo-ku, Tokyo 113-8656, Japan
}%
\author{Keitaro Anai}%
\affiliation{%
 Department of Applied Physics, School of Engineering, The University of Tokyo,\\
 7-3-1 Hongo, Bunkyo-ku, Tokyo 113-8656, Japan
}%
\author{Kenta Udagawa}
\affiliation{%
 Department of Applied Physics, School of Engineering, The University of Tokyo,\\
 7-3-1 Hongo, Bunkyo-ku, Tokyo 113-8656, Japan
}%
\author{Shuntaro Takeda}
\email{takeda@ap.t.u-tokyo.ac.jp}
\affiliation{%
 Department of Applied Physics, School of Engineering, The University of Tokyo,\\
 7-3-1 Hongo, Bunkyo-ku, Tokyo 113-8656, Japan
}%





\date{\today}
\begin{abstract}
\noindent 
Variational quantum algorithms (VQAs) provide a promising approach to achieving quantum advantage for practical problems on near-term noisy intermediate-scale quantum (NISQ) devices. Thus far, most studies on VQAs have focused on qubit-based systems, but the power of VQAs can be potentially boosted by exploiting infinite-dimensional continuous-variable (CV) systems. Here, we implement the CV version of one VQA, a quantum approximate optimization algorithm by developing an automated collaborative computing system between a programmable photonic quantum computer and a classical computer.
 We experimentally demonstrate that this algorithm solves the minimization problem of simple continuous functions by implementing the quantum version of gradient descent to localize an initially broadly-distributed wavefunction to the minimum. This method allows the execution of a practical CV quantum algorithm on a physical platform. Our work can be extended to the minimization of more general functions, providing an alternative to achieve the quantum advantage in practical problems.
\end{abstract}

\maketitle


\section{Introduction}
Variational quantum algorithms (VQAs) have recently emerged as the leading approach to achieving quantum advantage for practical problems under the constraints of near-term noisy intermediate-scale quantum (NISQ) devices \cite{preskill2018,cerezo2021}.
In VQAs, such constraints are avoided by the common strategy to repeatedly run shallow-depth quantum circuits with the circuit parameters updated by classical optimizers.
This strategy enables us to mitigate the accumulation of errors and fully exploit the computational space offered by the limited-scale devices.
Thus far, a wide variety of VQAs have been proposed theoretically for qubit-based systems, such as ones for combinatorial optimization \cite{farhi2014}, chemistry simulation \cite{peruzzo2014}, and machine learning \cite{mitarai2018}.
They have already been demonstrated experimentally on several physical platforms \cite{peruzzo2014,otterbach2017,bengtsson2020,pagano2020,harrigan2021,havlicek2019}.

In contrast, there have been much fewer proposals \cite{verdon2019,killoran2019,arrazola2019,volkoff2021,stechly2019} and no experimental implementations on continuous-variable (CV) VQAs, although CV quantum computing can potentially offer superior computational power in the NISQ era.
The potential of CV systems lies in the ability to process infinite-dimensional quantum information even on single-mode devices, while in qubit-based systems each qubit provides only two-dimensional computational space.
Furthermore, CV systems natively and efficiently handle continuous real parameters that often appear in real-world problems.
In general, fully exploiting such infinite dimensionality and continuous degree of freedom in CV systems for quantum computation has been regarded as impractical due to their noise sensitivity and difficulty in error correction \cite{takeda2019a}.
However, this can be in turn a promising approach to extracting high computational power in the NISQ era when the error correction is not assumed.

In this article, we implement the CV version \cite{verdon2019} of one of the most typical VQAs, a quantum approximate optimization algorithm (QAOA) \cite{farhi2014} by developing an automated collaborative computing system between a programmable photonic quantum computer and a classical computer.
We choose the simplest problem for the CV-QAOA and experimentally demonstrate it, where this algorithm minimizes one-variable continuous quadratic functions by working as the quantum version of gradient descent and localizing an initially broadly-distributed wavefunction to the minimum.
The algorithm is shown to robustly find approximate answers to the problem with a noisy shallow-depth quantum circuit, thus confirming that a VQA works also in CV systems.
This method showcases a CV quantum algorithm that can be 
used to solve practical problems, except for Gaussian boson 
sampling \cite{hamilton2017,zhong2021,madsen2022}, which is designed for achieving quantum supremacy and has been partially linked to some practical problems \cite{huh2015,arrazola2018}.
We also show that our implementation can be extended to the minimization problems of arbitrary-order functions by adding optical resources \cite{marek2018a}.
It can also be extended to multivariable functions by incorporating multi-mode interactions.
Thus our work highlights a new approach using CV quantum computing in the NISQ era, offering an alternative to achieve the quantum advantage in practical problems.

\section{Theory of CV-QAOA}
Many practical problems in various fields come down to optimization or minimization, which often require high computational costs for classical computers. The QAOA is a heuristic algorithm that could potentially offer a quantum speed-up to solve such problems on NISQ devices \cite{farhi2014}.
Theoretical aspects and experimental implementations of the QAOA have been recently studied intensively on qubit-based systems to solve discrete combinatorial optimization problems \cite{zhou2020,otterbach2017,willsch2020,bengtsson2020,pagano2020,harrigan2021}.

Later, a CV version of the QAOA was proposed to solve continuous optimization problems on CV systems \cite{verdon2019}. This algorithm is designed for minimizations of continuous real-valued functions, which have many practical applications in finance \cite{cornuejols2006}, machine learning \cite{aggarwal2020}, and engineering \cite{rao1996}. 
This proposal \cite{verdon2019} indicates that the CV-QAOA has potential of a quantum speed-up as its circuits can encode CV Grover's search algorithm \cite{pati2000}, which achieves a quadratic speed-up over the classical algorithms.

The CV-QAOA is formulated as follows.
The goal of the algorithm is to find an approximate minimum of a real-valued function $f(\bm{x})$ with $\bm{x}=(x_1,x_2,\cdots,x_N)\in \mathbb{R}^N$.
Let us consider a quantum-mechanical particle in the $N$-dimensional space with $[\hat{x}_i,\,\hat{p}_j]=i\delta_{ij}$, where $(\hat{x}_1,\hat{x}_2,\cdots,\hat{x}_N)=\hat{\bm{x}}$ and $(\hat{p}_1,\hat{p}_2,\cdots,\hat{p}_N)=\hat{\bm{p}}$ are position and momentum of the particle, respectively.
The initial state is $|\bm{p}=0\rangle$, which is an eigenstate of $\hat{\bm{p}}$ and thus equally-weighted superposition of $|\bm{x}\rangle$ for all $\bm{x}$.
The unitary operator given by
\begin{equation}
\hat{U}(\bm{\eta},\bm{\gamma})=\prod_{j=1}^{P}\mathrm{e}^{-i\gamma_j\hat{H}_M}\mathrm{e}^{-i\eta_j\hat{H}_C},
\end{equation}
where $\hat{H}_C=f(\hat{\bm{x}})$ and $\hat{H}_M=\hat{\bm{p}}^2/2$, transforms the initial state to the final state.
$P$ determines the circuit depth.
$\hat{H}_C$ and $\hat{H}_M$ are called cost and mixer Hamiltonians, respectively, and $\bm{\eta}=(\eta_1,\eta_2,\cdots,\eta_P)$ and $\bm{\gamma}=(\gamma_1,\gamma_2,\cdots,\gamma_P)$ are the tunable real positive parameters. 
The final state is then measured in the $\hat{\bm{x}}$-basis, the measurement outcome $\bm{x}$ being a candidate of the minimum of the function.
This algorithm can be considered as a quantum version of the gradient descent method;
a pair of the cost and mixer Hamiltonians transforms $\hat{\bm{x}}$ as
\begin{equation}
\hat{\bm{x}}\rightarrow\hat{\bm{x}}+\gamma_j\hat{\bm{p}}-\eta_j\gamma_j\nabla f(\hat{\bm{x}})\label{gradient_theory}
\end{equation}
and indeed $\hat{U}(\bm{\eta},\bm{\gamma})$ represents the Trotterized approximation of the time-evolution of a particle trapped in a potential of $f(\hat{\bm{x}})$ in an $N$-dimensional space \cite{farhi2014,verdon2019}.
Thus, while the distribution of $\bm{x}$ is initially uniform, it moves under the influence of the potential $f(\hat{\bm{x}})$ and then localizes around the minimum if the parameters $(\bm{\eta},\bm{\gamma})$ are properly chosen.

For the demonstration of the algorithm, we adopt the simplest problem setup, minimizing a quadratic function of one variable with the shallowest depth.
Specifically, we choose $f(x)=(x-a)^2$ with $a\in \mathbb{R}$ ($N=1$) and set $P=1$.

\begin{figure*}
\centering
\includegraphics[width=16cm]{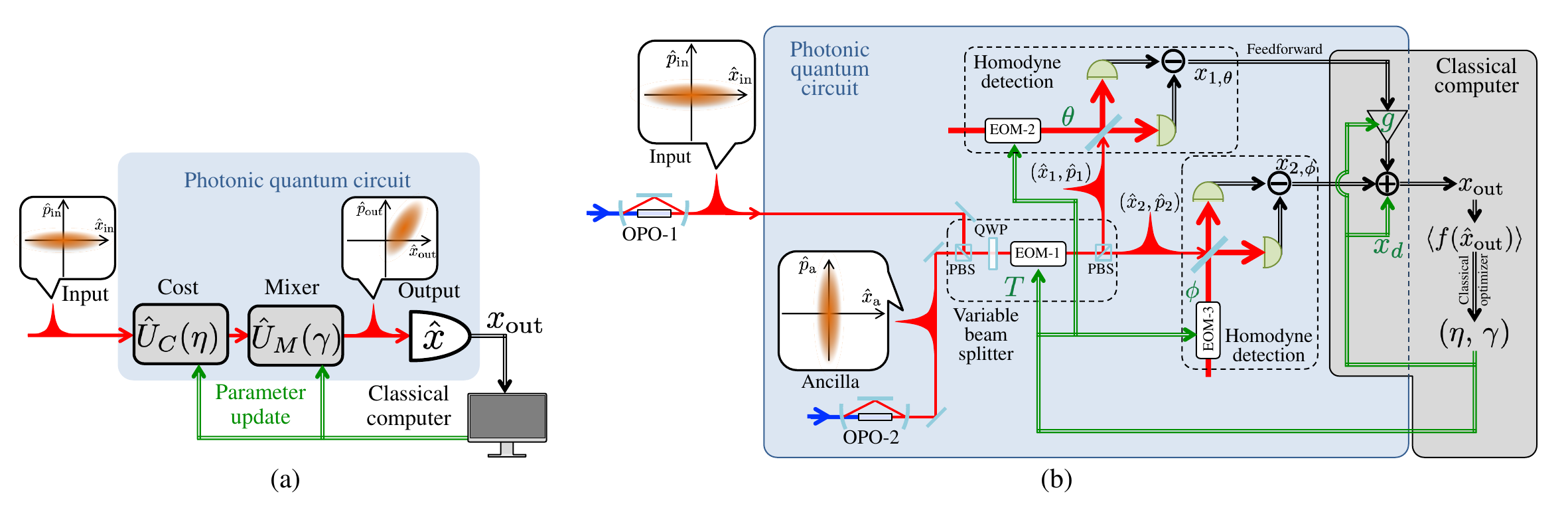}
\caption{Experimental implementation of the CV-QAOA. (a) Conceptual diagram of our demonstration. The input state is a squeezed state. The cost operator $\hat{U}_C(\eta)$ and the mixer operator $\hat{U}_M(\gamma)$ are applied to it, and the output state is measured in the $\hat{x}$-basis. The cost and mixer operators are updated following a certain protocol according to the circuit output $x_\mathrm{out}$. (b) Experimental setup. The input state is produced by OPO-1, and the ancillary state for the measurement-induced operation is produced by OPO-2. They interfere at the beam splitter having a variable transmissivity of $T$, and each of the two beams from the beam splitter is measured by the homodyne detector with a programmable measurement basis $\hat{x}_{1,\theta}$ and $\hat{x}_{2,\phi}$.
The feedforward operation proportional to $\hat{x}_{1,\theta}$ and the constant displacement operation are applied to $\hat{x}_{2,\phi}$ in a post process, which yields the circuit output $x_\mathrm{out}$.
The circuit parameters $(T,\,\theta,\,\phi)$, the feedforward gain $g$, and the constant displacement $x_d$ are determined according to a set of $(\eta,\,\gamma)$ so that the measurement-induced gate is performed in a consistent way. EOM, electro-optic modulator; PBS, polarizing beam splitter; QWP, quarter-wave plate.}
\label{concept_combined}
\end{figure*}

\vskip\baselineskip
\section{Photonic-circuit Implementation}
We implement the CV-QAOA on a programmable photonic quantum computer, where optical amplitude and phase are identified with position and momentum in the algorithm, respectively.
Our implementation is enabled by recent technological advances in photonic CV quantum computing, including state preparations, gate operations, and measurements \cite{fukui2022}.
Especially, programmable and multi-step CV gate operations have been demonstrated very recently, which are indispensable for implementing the CV-QAOA \cite{asavanant2021a,larsen2021a,enomoto2021}. Such previous studies have been limited to the proof-of-principle demonstrations of predetermined quantum gates. Here, we implement the CV-QAOA by developing an automated collaborative computing system between such a programmable CV photonic quantum computer and a classical computer, where the latter assesses the output of the former and automatically updates the program of the former in real time.

\subsection{Concept of our implementation}
Our implementation can be conceptually shown in Fig.~\ref{concept_combined}\blue{(a)}.
The input state is a $p$-squeezed state, which approximates $|p=0\rangle$.
After the parametrized operation of the cost and mixer on the input state, the output state is measured in the $\hat{x}$-basis.
According to Eq. (\ref{gradient_theory}), $\hat{U}(\eta,\gamma)$ in our specific problem setting transforms $\hat{x}$ as
\begin{equation}
\hat{x}_\mathrm{out} = (1-2\eta\gamma)\hat{x}_\mathrm{in}  + \gamma\hat{p}_\mathrm{in} + 2a\eta\gamma, \label{in-out_theory}
\end{equation}
$(\hat{x}_\mathrm{in},\hat{p}_\mathrm{in})$ and $(\hat{x}_\mathrm{out},\hat{p}_\mathrm{out})$ being the quadrature amplitude of the input and output states, respectively.
The parameters $\eta$ and $\gamma$ are iteratively updated according to the sampling results of the output state.

To realize the concept of Fig.~\ref{concept_combined}\blue{(a)}, we design and implement an optical and electrical setup shown in Fig.~\ref{concept_combined}\blue{(b)}.
To perform a gate operation of Eq.~(\ref{in-out_theory}) in a programmable way, we use a measurement-induced gate composed of an ancillary state, measurement, and feedforward. 
In fact, this setup can be regarded as the simplest version of a more general setup to implement the CV-QAOA for arbitrary-order functions \cite{marek2018a}, as we show in Appendix \ref{apb}.
The following is a specific description of the current setup.
The input state is produced by an optical parametric oscillator (OPO) named OPO-1.
The wavefunction of the state is squeezed in the $p$-direction in the sense that $\langle \hat{x}_\mathrm{in}^2\rangle=\mathrm{e}^{2r_+}/2$ and $\langle \hat{p}_\mathrm{in}^2\rangle=\mathrm{e}^{-2r_-}/2$ with $r_+,\,r_->0$.
The ancillary state from OPO-2 is an orthogonally squeezed state with $\langle \hat{x}_\mathrm{a}^2\rangle=\mathrm{e}^{-2r_-}/2$ and $\langle \hat{p}_\mathrm{a}^2\rangle=\mathrm{e}^{2r_+}/2$.
These two fields interfere at the beam splitter having variable transmissivity $T$ and then they are sent to two homodyne detectors.
The homodyne detectors measure the quadrature amplitudes $\hat{x}_{1,\theta} = \hat{x}_1\cos\theta+\hat{p}_1\sin\theta$ and $\hat{x}_{2,\phi} = \hat{x}_2\cos\phi+\hat{p}_2\sin\phi$.
Here $(\hat{x}_1,\hat{p}_1)$ and $(\hat{x}_2,\hat{p}_2)$ are the quadrature amplitudes of the upper and lower beams coming out of the variable beam splitter, respectively.
The parameters of the processor $T$, $\theta$, and $\phi$ can be varied via the applied voltage to the corresponding electro-optic modulators (EOMs).
Then $\hat{x}_{1,\theta}$ is fed forward to $\hat{x}_{2,\phi}$ with a certain gain.
To let the above-described circuit act as the gate designated by $(\eta,\gamma)$ as Eq.~(\ref{in-out_theory}), the parameters of the optical circuit are determined by $T=1/(1+\gamma^2)$, $\tan\theta=\gamma^2/[\gamma-2\eta(1+\gamma^2)]$, and $\tan\phi=\gamma$, and correspondingly the feedforward gain is set by $g=\sqrt{\gamma^2 - 4\eta\gamma + 4\eta^2(1+\gamma^2)}$, in which settings the contribution of the anti-squeezed quadrature of the ancilla $\hat{p}_\mathrm{a}$ to the gate output is canceled so that the measurement-induced gate operates properly.
The constant displacement of $x_d=2a\eta\gamma$ is also performed for the constant term in Eq.~(\ref{in-out_theory}).
The feedforward and constant displacing are numerically done by post-processing in the classical computer, which is justified because it gives the same results as those given by optical displacing.
In fact, feedforward operations were performed by post-processing in recent demonstrations of one-way quantum computation \cite{asavanant2021a,larsen2021a}.
After the above classical post-processing, the gate output $\hat{x}_\mathrm{out}$ becomes
\begin{equation}
\hat{x}_\mathrm{out} = (1-2\eta\gamma)\hat{x}_\mathrm{in}+\gamma\hat{p}_\mathrm{in}+2a\eta\gamma-2\eta\hat{x}_\mathrm{a},
\label{eq4}
\end{equation}
which asymptotically coincides with Eq.~(\ref{in-out_theory}) in the high squeezing limit $r_-\rightarrow\infty$ in the sense that the variance of the noise term $-2\eta\hat{x}_\mathrm{a}$ approaches zero (see Appendix \ref{apa} for the derivations).
Therefore, our photonic processor depicted in Fig.~\ref{concept_combined}\blue{(b)} is capable of obtaining the $x$-measurement result of the output state, $\hat{x}_\mathrm{out}$, with the gate parameters $\eta$ and $\gamma$ varied.

\begin{figure*}
\centering
\includegraphics[width=16cm]{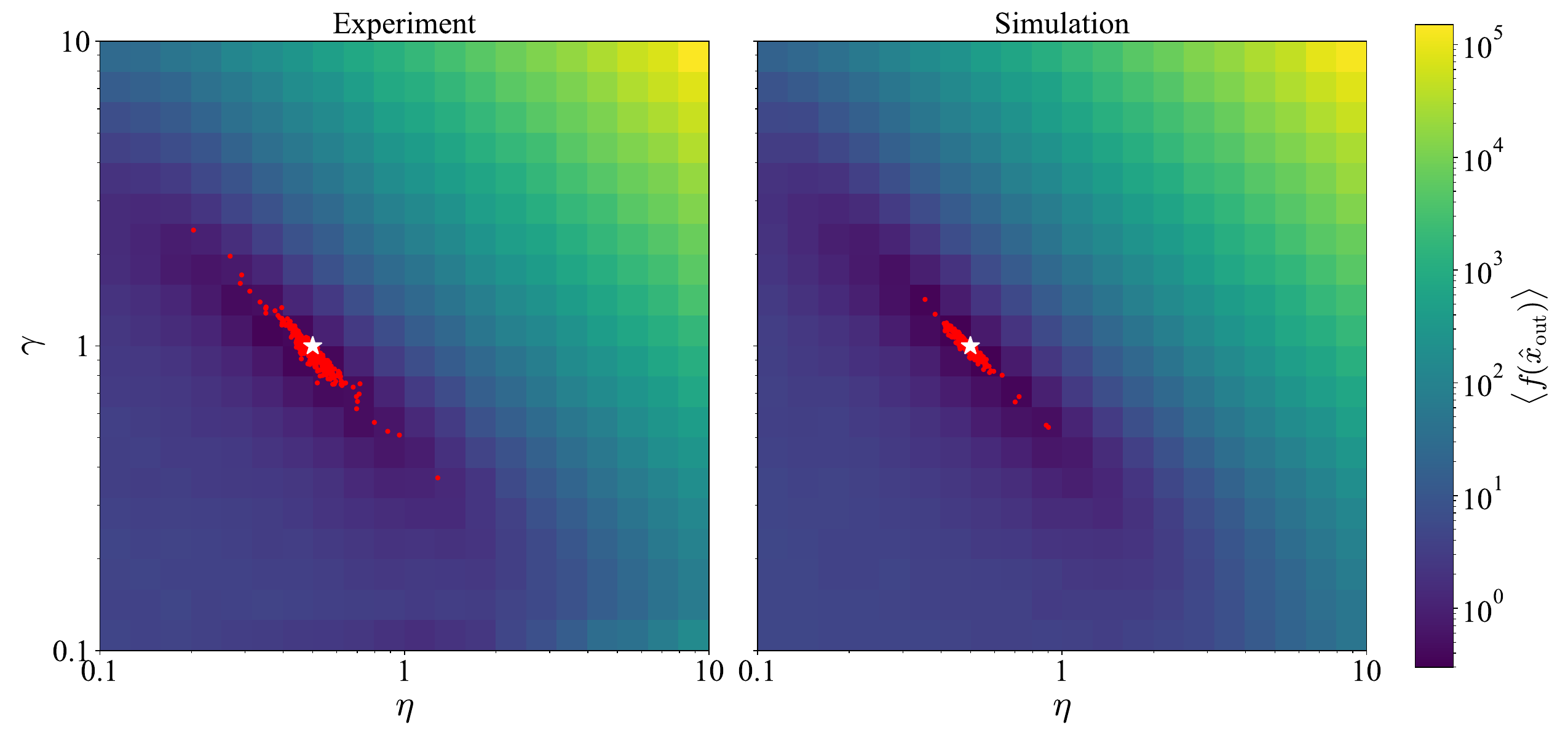}
\caption{Experimental and simulated landscapes of the CV-QAOA. The landscape structure of the experiment (left) agrees with that of the simulation (right). Each grid is evaluated by 1000 samples of $\hat{x}_\mathrm{out}$. $f(x)=(x-1)^2$ is used for the landscape evaluations. Each red dot on the landscapes shows the parameter set that records the smallest $\langle f(\hat{x}_\mathrm{out})\rangle$ during the 100-step Bayesian optimization. The white stars denote the theoretical optimum of $(\eta,\gamma)=(1/2,1).$ See Appendix \ref{ape} for the simulation condition.
}\label{landscape}
\end{figure*}

\subsection{Experimental setup}
Here we describe the details of our experimental setup in Fig.~\ref{concept_combined}\blue{(b)}. We use a continuous-wave laser of wavelength $1545.3\,\mathrm{nm}$.
Two OPOs are pumped by the second harmonic fields with wavelength of $772.7\,\mathrm{nm}$.
The pump power is set to $200\,\mathrm{mW}$.
The full width at the half-maximum of the OPOs is $60\,\mathrm{MHz}$.
The variable beam splitter is composed of a bulk electro-optic modulator named EOM-1, a quarter-wave plate, and a pair of polarizing beam splitters.
EOM-1 serves as a variable polarization rotator and thus works as a variable beam splitter with polarization optics.
We inserted the quarter-wave plate so that the transmissivity is $50\,\%$ when no voltage is applied to EOM-1, which makes it easy to lock the relative phase between the input and the ancillary beams.

Each homodyne detection is performed by interfering the local oscillator field with the signal field at a 50:50 beam splitter.
Two beams from the beam splitter are received by two photodiodes, the photocurrents of which are subtracted with each other and amplified in the electric circuit.
The bandwidth of the circuit is about $200\,\mathrm{MHz}$.
The optical power of the local oscillator field is set to $10\,\mathrm{mW}$.
A fiber-coupled electro-optic modulator EOM-2 or 3 shifts the optical phase of each local oscillator for the control of the homodyne angle $\theta$ or $\phi$.

The outcome of the homodyne detection is acquired by an oscilloscope and then sent to the classical computer.
The time series from the oscilloscope is converted to a quadrature amplitude by convoluting it with a mode function $h(t)$ defined by
\begin{equation}
h(t) = 
\begin{cases}
t\,\mathrm{e}^{-\Gamma^2t^2}&(|t|<t_1)\\
0&(\text{otherwise})
\end{cases},
\end{equation}
where $\Gamma=3\times10^7\,\mathrm{/s}$ and $t_1=50\,\mathrm{ns}$.
The purpose of using this mode function is to eliminate undesirable effect from low-frequency electrical noise from homodyne detectors \cite{yoshikawa2016}.
Based on the measured quadratures and the subsequent analysis, the classical computer can automatically reprogram the photonic quantum circuit by changing the voltages applied to EOM-1, 2, and 3. This enables the classical computer to collaborate with the photonic quantum computer updated in real time and perform the CV-QAOA.

As a preliminary measurement, the outputs of OPO-1 and OPO-2 are measured by the homodyne detectors with the transmissivity of the variable beam splitter set to zero.
The squeezing level and the anti-squeezing level of these modes are measured to be $-5.3\,\mathrm{dB}$ and $+9.0\,\mathrm{dB}$ on average.
This measurement result indicates that the overall optical loss of the experimental setup is estimated to be $22\,\%$.

\vskip\baselineskip
\section{Optimization landscape and algorithm performance}
The CV-QAOA is performed on the processor as follows;
first, we repeatedly run the circuit and sample $\hat{x}_\mathrm{out}$ with the parameters $\eta$ and $\gamma$ fixed to calculate the mean value $\langle f(\hat{x}_\mathrm{out}) \rangle$;
then, an outer-loop optimizer in the classical computer suggests new parameters to decrease $\langle f(\hat{x}_\mathrm{out}) \rangle$.
These two stages are repeated alternately.
To experimentally demonstrate the capability of the above, we operate the whole system in three different conditions for the parameter update:
(i) the parameters $\eta$ and $\gamma$ are scanned like a grid search;
(ii) the parameters are fixed at their optimum;
(iii) the parameters are updated according to the protocol of the Bayesian optimization.
Each result is shown in the following.

First, to qualitatively diagnose that the processor operates properly for the full search range of the outer-loop optimizer, $\langle f(\hat{x}_\mathrm{out}) \rangle$ is evaluated with the parameters $\eta$ and $\gamma$ scanned like a grid search.
The landscapes of $\langle f(\hat{x}_\mathrm{out}) \rangle$ as a function of $\eta$ and $\gamma$ are obtained from the experiment or the numerical simulation as shown in Fig.~\ref{landscape}.
The landscape structure of the experiment reasonably well agrees with that of the simulation.
Specifically, $\langle f(\hat{x}_\mathrm{out}) \rangle$ is small around $\eta\gamma=1/2$ and the smallest around $(\eta,\gamma)=(1/2,1)$, which is the theoretical optimum in the high squeezing limit indicated by white stars in the figure.
Note that the optimal point in the finite-squeezing case is $(\eta,\gamma)=(\sqrt{1-\delta}/2,\sqrt{1-\delta})$, where $\delta=\mathrm{e}^{-2r_-}/(\mathrm{e}^{2r_+}+2a^2)$, and thus the deviation from $(1/2,1)$ is less than $2\,\%$ with the values of $r_+$ and $r_-$ for our experiment no matter the value of $a$.
The overall landscape structure is also independent of the value of $a$ because it only affects the steepness of the valley of $\eta\gamma\sim1/2$.

\begin{figure}
\centering
\includegraphics[width=8cm]{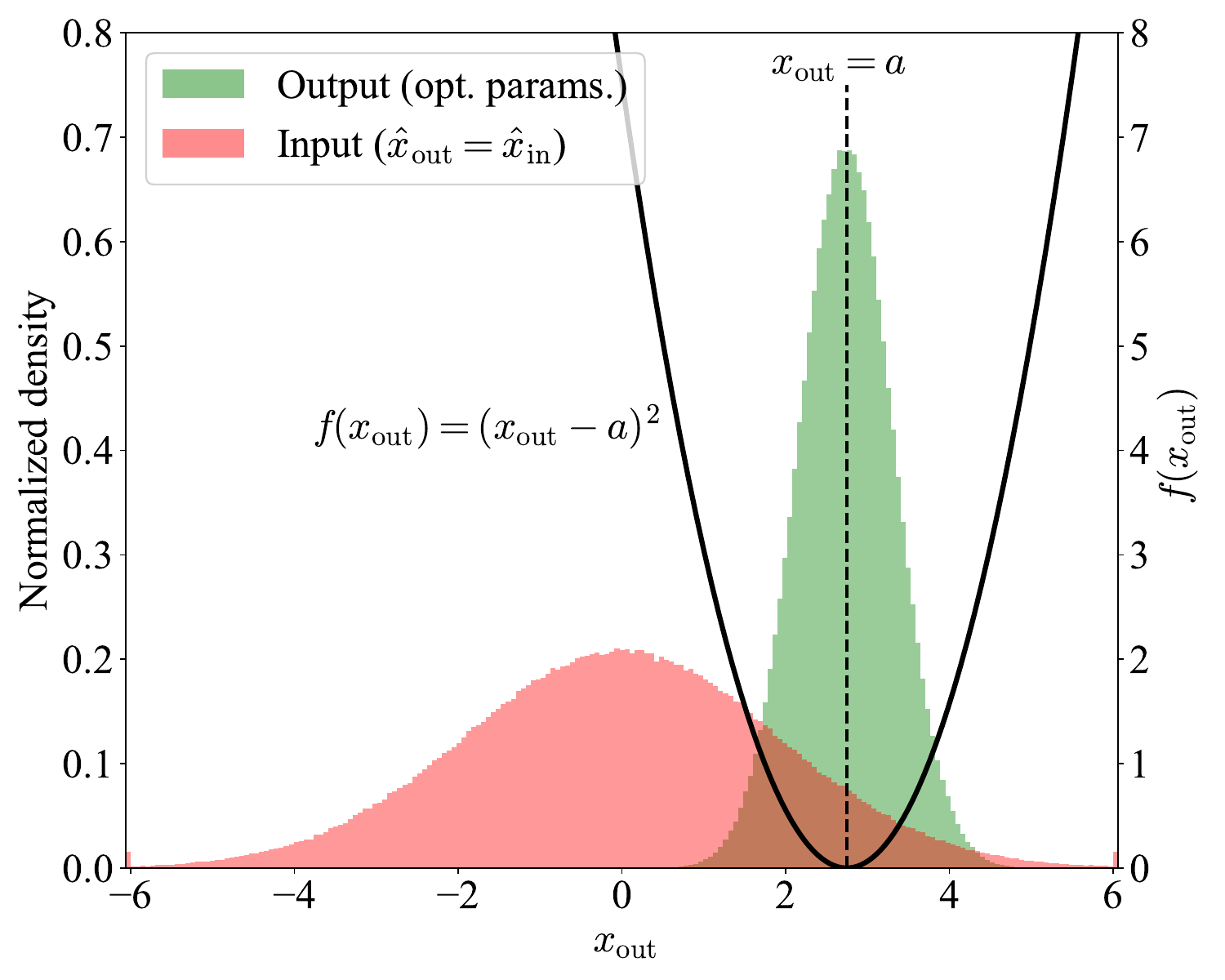}
\caption{Histogram of the output distribution. The normalized frequencies of the sampled $x_\mathrm{out}$ in the experiment are shown. The green bars are the output distribution with the optimal parameters of $(\eta,\gamma)=(1/2,1)$ while the red ones are the distribution of the input state, which corresponds to $(\eta,\gamma)=(0,0)$. The number of the sampled $x_\mathrm{out}$ is $1.1\times10^6$ for each. The curve of $f(x_\mathrm{out})=(x_\mathrm{out}-a)^2$ is overlaid. We use $a=2.745$ for these measurements. The comparison of two distributions demonstrates how the CV-QAOA works.}\label{gdm}
\end{figure}

Next, to visualize how the algorithm works like the gradient descent method to reshape the wavefunction of $\hat{x}$, we run the processor with the parameters fixed at the optimal point $(\eta,\gamma)=(1/2,1)$.
Figure \ref{gdm} shows the histogram of the sampled $\hat{x}_\mathrm{out}$ for a specific $a$.
The sampling results of the input state $\hat{x}_\mathrm{in}$ are also plotted.
It can be found that the input state has a broad distribution, and the optimal gate operation localizes the distribution around $a$, the exact solution.
In fact, the standard deviation of $\hat{x}_\mathrm{out}$ is $0.5767(4)$, smaller than that of the vacuum state $1/\sqrt{2}$ thanks to the quantumness of the processor.
Also, the gap between the mean output and the exact solution is $\langle \hat{x}_\mathrm{out}\rangle-a=-7.1(5)\times 10^{-3}$, the absolute value of which is much smaller than the standard deviation.
This indicates that the systematic shift from the exact solution is negligible with respect to the statistical broadening of the distribution.
These results visually prove that the experimentally implemented gate for $\hat{U}(\eta,\gamma)$ in our problem setting properly localizes the distribution around the minimum like the gradient descent method when the parameters are optimal.

\begin{figure*}
\centering
\includegraphics[width=17cm]{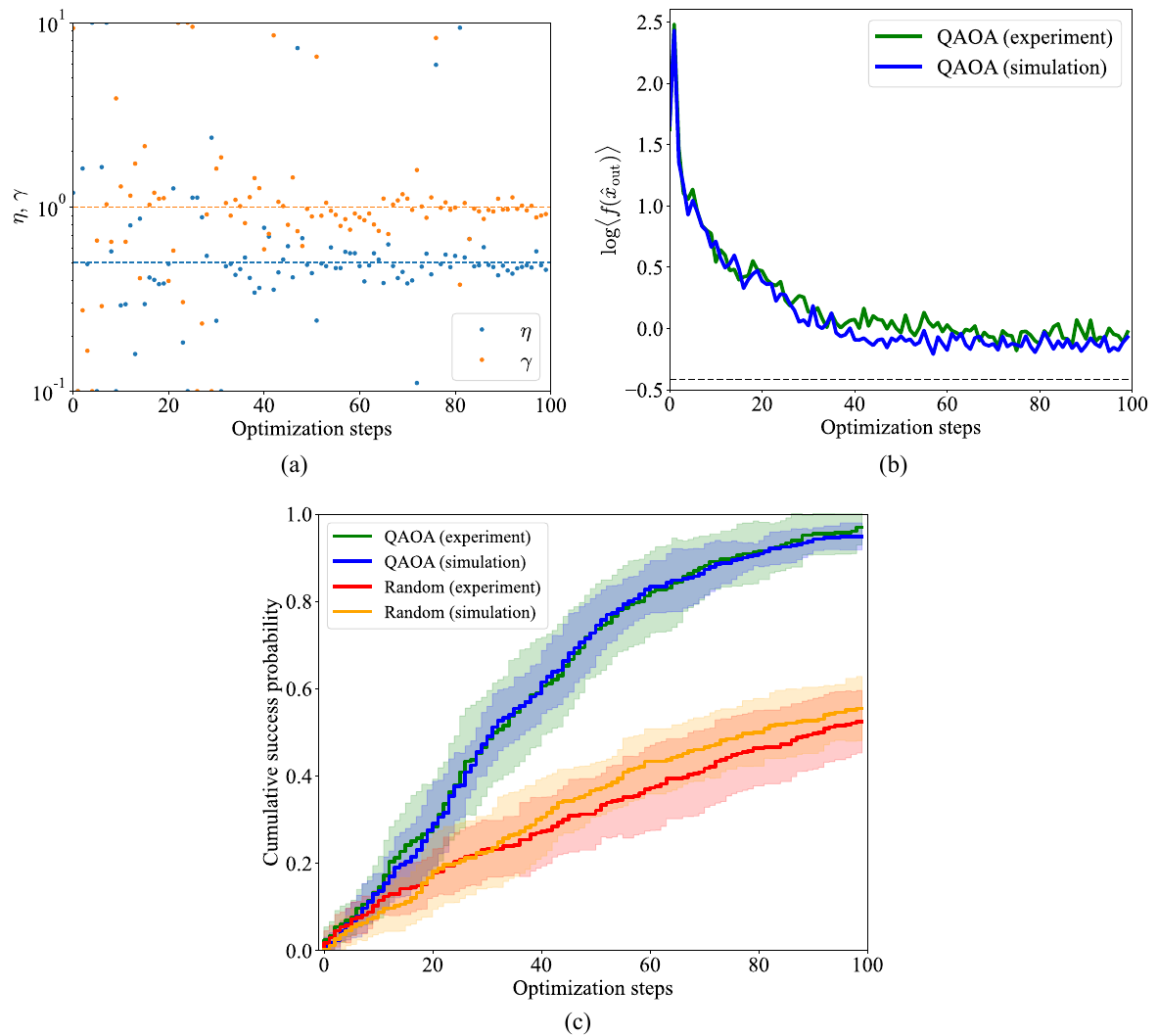}
\caption{Algorithm performance with parameter updates. (a) Typical trace of the parameters updated by the Bayesian optimization. The initial search uniformly spans the full range but gradually the search becomes concentrated around the optimum, which is depicted by the dashed lines. (b) Convergence of the classical optimization. The solid traces show the average of the target function $\ln\langle f(\hat{x}_\mathrm{out})\rangle$ from the experiment and the simulation. The black dashed line is the theoretical values for the case where the parameters are fixed at $(\eta,\,\gamma)=(1/2,\,1)$. (c) Success probability to sample $x_\mathrm{out}$ such that $f(x_\mathrm{out})<1\times10^{-9}$. The solid line is derived by averaging the probability for the eleven sets of such trials. The shaded area shows the $\pm1\sigma$ region around the average. The green and blue plots are the experimental and simulated results of the CV-QAOA, respectively. As a reference, the red and orange plots show the experimental and simulated results of random sampling. They indicate that the CV-QAOA finds the minimum of $f(x)$ significantly more efficiently than the random sampling.}
\label{results_combined}
\end{figure*}

Finally, to evaluate the performance of the CV-QAOA in a realistic condition where the optimum of the parameters is unknown, we perform the algorithm with the parameter updated by the Bayesian optimization.
The optimizer suggests new parameters every 1000 samples of $\hat{x}_\mathrm{out}$ so that $\ln\langle f(\hat{x}_\mathrm{out}) \rangle$ is minimized (see Appendix \ref{apc}).
We repeat such a process many times with the value of $a$ changed to see the statistical behavior of the algorithm (see Appendix \ref{apd}).
In Fig.~\ref{landscape}, the overlaid red dots show the distribution of the classically optimized pair of $(\eta,\gamma)$.
Each dot corresponds to the pair of the parameters that gives the smallest $\langle f(\hat{x}_\mathrm{out}) \rangle$ among 100 suggestions by the optimizer for each execution of the CV-QAOA.
These figures show that the classical optimizer reaches the point around the optimum within the 100 steps as the simulation predicts.
Figure \ref{results_combined}\blue{(a)} shows the typical trace of the parameter update by the Bayesian optimization, where a wide area is initially explored but gradually the search becomes concentrated around the optimum.
The behavior of $\ln\langle f(\hat{x}_\mathrm{out}) \rangle$, which is the target function of the Bayesian optimization, is shown in Fig.~\ref{results_combined}\blue{(b)}.
The traces denote the average of $\ln\langle f(\hat{x}_\mathrm{out}) \rangle$ for all the CV-QAOA trials.
The observed decreasing trend of the average of the target function is comparable to the simulation expectation.
The small differences between the experimental and simulated traces can be attributed to imperfections of the optical setup such as intensity fluctuations and/or alignment drifts of the local oscillator beams for the homodyne detectors.

To quantify the performance in minimizing the function, the success probability of finding the minimum is evaluated experimentally and numerically.
Figure~\ref{results_combined}\blue{(c)} shows the cumulative success probability of finding the minimum of $f(x)$ up to a certain step of the parameter updates.
Here we set the criterion of the success by $f(x_\mathrm{out})=(x_\mathrm{out}-a)^2<1\times10^{-9}$.
The success probability is calculated by the success frequency for 30 different values of $a$ in $f(x)=(x-a)^2$.
The solid line is derived by averaging the success probability for eleven sets of the trials while the shaded area denotes the $\pm 1\sigma$-region around the average derived from the eleven trials.
These results show that the experimental results of the increase of the cumulative success probability coincide with the numerically simulated ones.
They also show that the success probability obtained by the CV-QAOA is significantly better than that by random sampling, where the input state is directly measured.

\section{Discussion}
In conclusion, we demonstrate the successful implementation of the CV-QAOA with the parameters of the programmable photonic quantum processor updated.
The demonstration experimentally shows that the performance of the algorithm for the minimization of quadratic functions is significantly better than that of the random sampling and comparable to what the numerical simulation predicts.
Even though the experimental system is influenced by various imperfections including optical loss, the CV-QAOA still successfully finds the minimum of the function, indicating that the algorithm works robustly.

Such imperfections limit the effective squeezing levels. The optimum effective squeezing level for the overall performance is nontrivial. In fact, with higher squeezing levels, the width of the output distribution in Fig.~\ref{gdm} gets narrower and this increases the probability of finding the optimal point of $x$. On the other hand, it is also observed in our numerical simulation that the parameters $\eta$ and $\gamma$ are optimized more slowly with higher squeezing levels due to the landscape change in Fig.~\ref{landscape}. The former (the latter) is advantageous (disadvantageous) to the algorithm. Optimum squeezing level should be investigated, but its detailed analysis is left for future work.

Our experiment in this paper can be simulated efficiently with classical computers because our system is entirely built with Gaussian building blocks~\cite{ths:ClassicalSimulation}. However, we show that our implementation can be extended to arbitrary-order functions by adding non-Gaussian ancillary states other than squeezed states \cite{marek2018a} (Appendix \ref{apb}). This may push the CV-QAOA to the non-Gaussian regime beyond efficient classical simulation.
.
It can also be extended to multivariable functions by using beam splitters for multi-mode interactions.
In this way, more complex functions can be minimized to address the practical problems.
For such more complex functions, we may set $P>1$ with a deeper circuit at the cost of the increased number of the classical parameters to be optimized $(\bm{\eta},\bm{\gamma})$.
The current system took $\sim$100 seconds to perform one CV-QAOA trial with 100 optimization steps. This runtime was mainly dominated by the measurement time, which can be shortened by increasing the bandwidths of the OPOs and the homodyne detectors.

{
This work is the first experimental realization of a quantum algorithm using CV information, which proves the usefulness of CV quantum systems for natively solving CV problems. In general, CV problems can also be solved with qubit-based quantum computers~\cite{ths:DVforCV}, but it requires more resources to represent discretized CV parameters with many qubits and also introduces quantization errors. CV quantum computers can avoid such difficulties and efficiently encode CV problems.
This demonstration sheds light on the advantages of CV quantum computing that its infinite dimensional space can be exploited in NISQ applications.
It stimulates the realizations of other VQAs in CV systems such as quantum machine learning \cite{killoran2019,arrazola2019} and thus opens a new promising way toward quantum advantage.
}

\section*{Acknowledgments}
The authors thank Keisuke Fujii, Kosuke Fukui, and Kosuke Mitarai for valuable discussions. This work was partly supported by JSPS KAKENHI Grant Numbers 20H01833 and 21K18593, MEXT Leading Initiative for Excellent Young Researchers, Toray Science Foundation (19-6006), and the Canon Foundation.

\appendix

\section{INPUT-OUTPUT RELATION OF THE OPTICAL CIRCUIT}\label{apa}
\begin{figure}
\centering
\includegraphics[width=8cm]{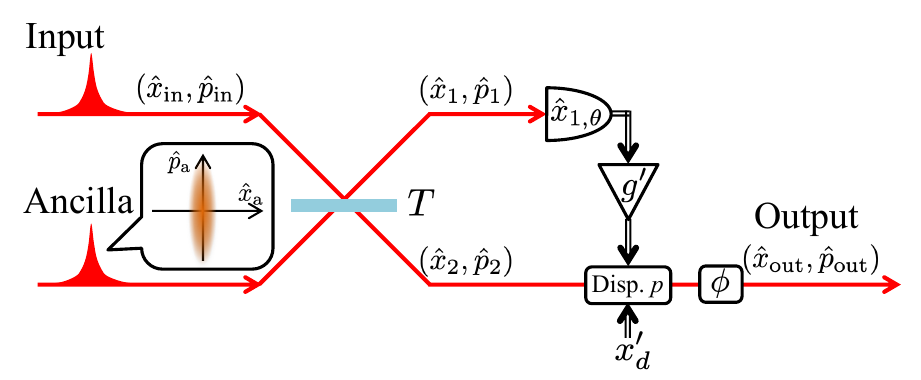}
\caption{Optical circuit for a measurement-induced linear transformation. The input state interferes with the ancilla at the beamsplitter having the transmissivity of $T$. One outgoing beam from the beamsplitter is measured by homodyne detection, and the measurement outcome is fed forward to the other beam. The quadrature $\hat{p}_2$ is displaced by $g'\hat{x}_{1,\theta} + x'_d$, and then a phase rotation by $\phi$ is applied. }\label{S1}
\end{figure}

Let us describe why the optical circuit in Fig.~\ref{concept_combined}\blue{(b)} works as the unitary operation $\hat{U}(\eta,\gamma)=\mathrm{e}^{-i\gamma \hat{p}^2/2}\mathrm{e}^{-i\eta (\hat{x}-a)^2}$ for our problem setting and the successive measurement in $\hat{x}$-basis.
Here we recall that this unitary transforms the input state represented by $(\hat{x}_\mathrm{in},\hat{p}_\mathrm{in})$ into the output state represented by $(\hat{x}_\mathrm{out},\hat{p}_\mathrm{out})$ as
\begin{equation}
\begin{bmatrix}
\hat{x}_\mathrm{out}\\
\hat{p}_\mathrm{out}\\
\end{bmatrix}
=
\begin{bmatrix}
1-2\eta\gamma&\gamma\\
-2\eta&1\\
\end{bmatrix}
\begin{bmatrix}
\hat{x}_\mathrm{in}\\
\hat{p}_\mathrm{in}\\
\end{bmatrix}
+
\begin{bmatrix}
2a\eta\gamma\\
2a\eta\\
\end{bmatrix}\label{in-out_theory_SI}.
\end{equation}

One of the possible implementations of such a linear transformation is shown in Fig.~\ref{S1}, which is based on the squeezing gate in Ref. \cite{miyata2014}.
First, we explain how this measurement-induced implementation works.
The ancillary state represented by $(\hat{x}_\mathrm{a},\hat{p}_\mathrm{a})$ is an $x$-squeezed state.
The output of this circuit is calculated as
\begin{align}
\begin{bmatrix}
\hat{x}_\mathrm{out}\\
\hat{p}_\mathrm{out}\\
\end{bmatrix}
&=
\begin{bmatrix}
\cos\phi&\sin\phi\\
-\sin\phi&\cos\phi\
\end{bmatrix}
\begin{bmatrix}
\hat{x}_2\\
\hat{p}_2+g'\hat{x}_{1,\theta}+x'_d\\
\end{bmatrix}\\
&=
\begin{bmatrix}
\hat{x}_{2,\phi}+g'\hat{x}_{1,\theta}\sin\phi+x'_d\sin\phi\\
\hat{p}_{2,\phi}+g'\hat{x}_{1,\theta}\cos\phi+x'_d\cos\phi\\
\end{bmatrix},\label{tochu}
\end{align}
where $g'$ is the feedforward gain and $x'_d$ is the constant displacing.
Here, $(\hat{x}_i,\hat{p}_i)$ and $(\hat{x}_{i,\psi},\hat{p}_{i,\psi})$ ($i=1,2$, and $\psi\in \mathbb{R}$) denote the quadrature amplitudes of the beams coming out of the beam splitter defined by
\begin{align}
\hat{x}_1&=\sqrt{1-T}\hat{x}_\mathrm{in}+\sqrt{T}\hat{x}_\mathrm{a},\\
\hat{p}_1&=\sqrt{1-T}\hat{p}_\mathrm{in}+\sqrt{T}\hat{p}_\mathrm{a},\\
\hat{x}_2&=\sqrt{T}\hat{x}_\mathrm{in}-\sqrt{1-T}\hat{x}_\mathrm{a},\\
\hat{p}_2&=\sqrt{T}\hat{p}_\mathrm{in}-\sqrt{1-T}\hat{p}_\mathrm{a},\\
\begin{bmatrix}
\hat{x}_{i,\psi}\\
\hat{p}_{i,\psi}
\end{bmatrix}
&=
\begin{bmatrix}
\cos\psi&\sin\psi\\
-\sin\psi&\cos\psi\
\end{bmatrix}
\begin{bmatrix}
\hat{x}_i\\
\hat{p}_i
\end{bmatrix}.
\end{align}
If we set the feedforward gain by $g'= \sqrt{(1-T)/T}\,/\sin\theta$, the output described by Eq.~(\ref{tochu}) becomes
\begin{equation}
\begin{split}
\begin{bmatrix}
\hat{x}_\mathrm{out}\\
\hat{p}_\mathrm{out}\\
\end{bmatrix}
&=\frac{1}{\sqrt{T}}
\begin{bmatrix}
(1-T)\cot\theta\sin\phi+T\cos\phi &  \sin\phi \\
(1-T)\cot\theta\cos\phi-T\sin\phi &  \cos\phi
\end{bmatrix}
\begin{bmatrix}
\hat{x}_\mathrm{in}\\
\hat{p}_\mathrm{in} 
\end{bmatrix}\\
&+
\begin{bmatrix}
\sin\phi\\
\cos\phi
\end{bmatrix}x'_d
+\sqrt{1-T}
\begin{bmatrix}
\cot\theta\sin\phi-\cos\phi \\
\cot\theta\cos\phi+\sin\phi 
\end{bmatrix}\hat{x}_\mathrm{a}.
\label{T_SI}
\end{split}
\end{equation}
Note that $g'$ is chosen so that the anti-squeezed quadrature $\hat{p}_a$ disappears in this expression.
This expression indicates that, apart from a noise term proportional to the squeezed quadrature $\hat{x}_\mathrm{a}$, the circuit in Fig.~\ref{S1} is capable of performing various linear transformations by changing $T$, $\theta$, $\phi$, and $x'_d$.
In fact, by setting $T=1/(1+\gamma^2)$, $\tan\theta=\gamma^2/[\gamma-2\eta(1+\gamma^2)]\,\,(0\le\theta<\pi)$, $\tan\phi=\gamma\,\,(0\le\phi<\pi/2)$, and $x'_d=2a\eta\sqrt{1+\gamma^2}$, Eq.~(\ref{T_SI}) becomes
\begin{equation}
\begin{bmatrix}
\hat{x}_\mathrm{out}\\
\hat{p}_\mathrm{out}\\
\end{bmatrix}
=
\begin{bmatrix}
1-2\eta\gamma&\gamma\\
-2\eta&1\\
\end{bmatrix}
\begin{bmatrix}
\hat{x}_\mathrm{in}\\
\hat{p}_\mathrm{in}\\
\end{bmatrix}
+
\begin{bmatrix}
2a\eta\gamma\\
2a\eta\\
\end{bmatrix}
+
\begin{bmatrix}
-2\eta\\
(\gamma-2\eta)/\gamma\\
\end{bmatrix}\hat{x}_\mathrm{a},
\end{equation}
which asymptotically coincides with Eq.~(\ref{in-out_theory_SI}) in the high squeezing limit of $\hat{x}_a\rightarrow0$.
Note that the upper row of this equation is identical to Eq.~(\ref{eq4}).

Given that we eventually measure the quadrature amplitude $\hat{x}_\mathrm{out}$, the phase rotation by $\phi$ can be achieved instead by changing the homodyne angle for the measurement of the output state.
In addition, the displacing operation can be replaced by numerical post-processing after the homodyne measurement as the displacing operation only shifts the mean value of the measurement outcome. 
For this reason, the circuit in Fig.~\ref{concept_combined}\blue{(b)} is equivalent to the one in Fig.~\ref{S1} as long as the quadrature amplitude of the output state is measured.
In the circuit in Fig.~\ref{concept_combined}\blue{(b)}, $\hat{x}_\mathrm{out}$ is in fact provided using the two homodyne-measurement outcomes by $\hat{x}_{2,\phi}+g\hat{x}_{1,\theta}+x_d$, which corresponds to the upper row of Eq.~(\ref{tochu}).
Here we redefined $g:=g'\sin\phi=\sqrt{\gamma^2-4\eta\gamma+4\eta^2(1+\gamma^2)}$ and $x_d:=x'_d\sin\phi=2a\eta\gamma$.

\section{IMPLEMENTATION OF HIGHER-ORDER FUNCTIONS}\label{apb}
\begin{figure*}
\centering
\includegraphics[width=12.5cm]{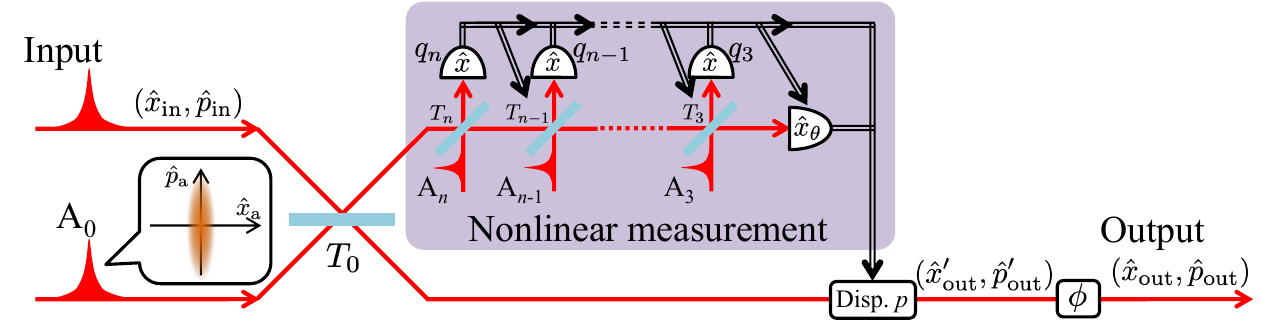}
\caption{Implementation of a higher-order function. The outgoing beam from the first beam splitter is nonlinearly measured with the help of the ancillary nonlinear phase states, and the measurement outcomes are fed forward before the phase rotation by $\phi$. Instead of also preparing a quadratic phase state and an additional beam splitter at the right end of nonlinear measurement block as proposed in Fig.~1d in Ref \cite{marek2018a}, the homodyne angle $\theta$ of the rightmost homodyne measurement is made variable. In fact, the implementation of the fourth-order nonlinear phase gate is proposed in this way (Fig.~2 in Ref. \cite{marek2018a}).}\label{n-th}
\end{figure*}
Let us show that our setup can be extended to a quantum circuit for minimizing a higher-order function.
As can be seen from Eq.~(\ref{gradient_theory}), if the minimized function $f(x)$ is an $n$-th order polynomial, the transformation by a pair of the cost and mixer Hamiltonians in the QAOA involves an $(n-1)$-th order polynomial of $\hat{x}$:
\begin{align}
\hat{x}&\rightarrow\hat{x}+\gamma\hat{p}-\eta\gamma \sum_{k=1}^n ka_k\hat{x}^{k-1},\label{n-th_x}\\
\hat{p}&\rightarrow\hat{p}-\eta \sum_{k=1}^n ka_k\hat{x}^{k-1},\label{n-th_p}
\end{align}
where we express $f(x)=\sum_{k=0}^{n}a_kx^k$ with real coefficients $a_k$.
The above transformation can be implemented by the circuit shown in Fig. \ref{n-th}.
In contrast to Fig.~\ref{S1}, where one beam from the beam splitter is measured by simple homodyne detection, the beam is nonlinearly measured with the help of the ancillary states.
The working principle can be made clear by dividing the circuit into two part: the nonlinear measurement followed by the feedforward, and the phase rotation by $\phi$.

The combination of the nonlinear measurement and feedforward provides an $(n-1)$-th order polynomial of $\hat{x}$ in the following way.
The ancillary states $\mathrm{A}_k$ ($3\le k\le n$) are $k$-th order phase states with reduced fluctuations in quadratures $\hat{p}_k-k\chi_k\hat{x}_k^{k-1}$, where $(\hat{x}_k,\hat{p}_k)$ are canonical pairs of the quadrature operators of the corresponding modes.
The beam is sequentially coupled with the ancillary states $\mathrm{A}_k$ by the beam splitters having transmissivity of $T_k$.
Then, one outgoing mode from each beam splitter is measured by homodyne detection in $\hat{x}$ quadrature, the outcome of which is labeled as $q_k$.
The transmissivity $T_k$ is adaptively controlled depending on the past outcomes $q_j$, where $k<j\le n$.
The rightmost homodyne measurement is done with the homodyne angle of $\theta$, which is also controlled depending on the past outcomes.
The quadratures after appropriate feedforwards from the nonlinear measurement block are expressed, in the limit that the ancillae are ideal, as
\begin{align}
\hat{x}'_\mathrm{out} &= \sqrt{T_0}\hat{x}_\mathrm{in},\\
\hat{p}'_\mathrm{out} &= \frac{1}{\sqrt{T_0}}\hat{p}_\mathrm{in}+\sum_{k=1}^nkC_k\hat{x}_\mathrm{in}^{k-1}.
\end{align}
where $C_k$ depends on $q_j$ ($k<j\le n$), $T_l$ ($k\le l\le n$) and $\theta$ (see Eqs.(5) and (9) in Ref. \cite{marek2018a}).
Here, by recurrently determining $T_n,\,T_{n-1},\cdots, T_3$ in the order from $n$ to $3$ and then $\theta$, the coefficient $C_k$ $(2\le k \le n)$ can be arbitrarily chosen.
As the coefficient $C_1$ corresponds to constant displacing for $\hat{p}$, it is also arbitrarily determined at the displacing operation.
Therefore, the above process realizes $n$-th order nonlinearity with arbitrary coefficients $C_k$.

The phase rotation by $\phi$ transforms $(\hat{x}'_\mathrm{out},\hat{p}'_\mathrm{out})$ as
\begin{align}
\hat{x}_\mathrm{out} &=\sqrt{T_0} \hat{x}_\mathrm{in}\cos\phi +\frac{\hat{p}_\mathrm{in}}{\sqrt{T_0}} \sin\phi +\sin\phi \sum_{k=1}^nkC_k\hat{x}_\mathrm{in}^{k-1},\\
\hat{p}_\mathrm{out} &=-\sqrt{T_0}\hat{x}_\mathrm{in}\sin\phi +\frac{\hat{p}_\mathrm{in}}{\sqrt{T_0}} \cos\phi +\cos\phi \sum_{k=1}^nkC_k\hat{x}_\mathrm{in}^{k-1}.
\end{align}
Here, by setting $\tan\phi=\gamma$, $T_0 = 1/(1+\gamma^2)$, $C_2=[\gamma/2- a_2\eta(1+\gamma^2)]/\sqrt{1+\gamma^2}$, and $C_k = -a_k\eta\sqrt{1+\gamma^2}$ $(k\ne 2)$, these equations become
\begin{align}
\hat{x}_\mathrm{out}&=\hat{x}_\mathrm{in}+\gamma\hat{p}_\mathrm{in}-\eta\gamma\sum_{k=1}^n ka_kx_\mathrm{in}^{k-1},\\
\hat{p}_\mathrm{out}&=\hat{p}_\mathrm{in}-\eta\sum_{k=1}^n ka_kx_\mathrm{in}^{k-1},
\end{align}
which realize the transformation of Eqs.~(\ref{n-th_x}) and (\ref{n-th_p}).
In this way, the transformation by a pair of the cost and mixer Hamiltonians for a given polynomial $f(x)$ can be implemented.
Thus, in the viewpoint of implementing generic higher-order functions, our experimental setup can be regarded as the simplest case of Fig.~\ref{n-th}.

\section{PARAMETER SEARCH CONDITION WITH THE BAYESIAN OPTIMIZATION}\label{apc}
For the update of the circuit parameters $(\eta,\gamma)$ in Fig.~\ref{concept_combined}, we adopt the Bayesian optimization, which has been commonly used among the derivative-free optimization methods in the parameter optimization of the QAOA \cite{otterbach2017,bengtsson2020}.
The following conditions of the Bayesian optimization are set after checking the convergence of the circuit parameters in numerical simulations.
$\log_{10}\eta$ and $\log_{10}\gamma$ are handed to the Bayesian optimizer as free parameters.
The target function to be optimized is $\ln\langle f(\hat{x}_\mathrm{out})\rangle$.
We use a python package for the Bayesian optimization \cite{package}.
The Mat\'ern kernel with $\nu=2.5$ is chosen as the covariance kernel function.
The acquisition function is a type called the upper confidence bound given by $\mu+\kappa(t)\sigma$, where $\mu$ and $\sigma$ are the estimated mean and standard deviation of the target function, respectively.
We set $\kappa(t)=\kappa_0\times0.97^t$, where $\kappa_0=2.576$ and $t$ is the number of optimization steps.

As for the QAOA of the qubit system, the parameter range of the unitary operations is often limited by $[0,\,2\pi)$.
However, the CV case does not have such a limit.
Since it is difficult to optimize parameters with unlimited search range, we limit the range by estimating the order of the optimum of the parameters using a generic prescription described below.
The search range of the parameters is set by $0.1<\eta<10$ and $0.1<\gamma<10$ because the optimum is estimated to be $(\eta_\mathrm{opt},\gamma_\mathrm{opt})\sim(1,1)$.

Here we explain the prescription for the estimation.
Let us consider the CV-QAOA with $P=1$ for a one-variable function $f(x)$.
If implemented by a combination of measurement-induced gates \cite{marek2018a}, a pair of the cost and mixer operations transforms the quadratures as
\begin{equation}
\hat{x}_\mathrm{out} = \hat{x}_\mathrm{in}+\gamma\hat{p}_\mathrm{in}-\eta\gamma\nabla f(\hat{x}_\mathrm{in})+N(\eta,\gamma;\hat{\bm{x}}_\mathrm{a},\hat{\bm{p}}_\mathrm{a}),\label{x_out}
\end{equation}
where $N$ is a noise term arising from non-ideal ancillary states.
$\hat{\bm{x}}_\mathrm{a}$ and $\hat{\bm{p}}_\mathrm{a}$ formally denote the quadrature amplitudes of the ancillary states. 
Let us assume that the ancillary states are linearly or non-linearly squeezed sufficiently and thus $|\langle N\rangle|\ll |x_\mathrm{min}|$, where $x_\mathrm{min} = \mathrm{argmin}f(x)$.
We can also assume that the optimal parameters, namely $(\eta_\mathrm{opt},\gamma_\mathrm{opt})$, localize the distribution of $\hat{x}_\mathrm{out}$ around $x_\mathrm{min}$, and thus provide two conditions: $\langle \hat{x}_\mathrm{out}\rangle\sim x_\mathrm{min}$ and $\langle\Delta\hat{x}_\mathrm{out}^2\rangle$ is minimized.
By assuming that the optimal parameters $(\eta_\mathrm{opt},\gamma_\mathrm{opt})$ are adopted and taking the expectation value of Eq.~(\ref{x_out}), the assumption $\langle \hat{x}_\mathrm{out}\rangle\sim x_\mathrm{min}$ reduces to
\begin{equation}
\eta_\mathrm{opt}\gamma_\mathrm{opt}\sim-\frac{x_\mathrm{min}}{\langle \nabla f(\hat{x}_\mathrm{in})\rangle}.\label{product}
\end{equation}
Here we assume $\langle\hat{x}_\mathrm{in}\rangle=0$ and $\langle\hat{p}_\mathrm{in}\rangle=0$ since the input state is the squeezed vacuum.
In this way, if the prior information on the order of magnitude of $x_\mathrm{min}$ and $\langle \nabla f(\hat{x}_\mathrm{in})\rangle$ is available, the product $\eta_\mathrm{opt}\gamma_\mathrm{opt}$ can be inferred.
Let us next consider the variance of $\hat{x}_\mathrm{out}$.
Since Eq.~(\ref{product}) estimates the product of the optimal parameters, we evaluate the variance under the constraint of $\eta\gamma=c$, where $c$ denotes the estimated value of the product $\eta_\mathrm{opt}\gamma_\mathrm{opt}$.
By substituting $\eta\gamma=c$ into Eq.~(\ref{x_out}) to eliminate $\eta$, we have
\begin{equation}
\begin{split}
\langle\Delta\hat{x}_\mathrm{out}^2\rangle&\sim \left\langle \Delta\left[ \hat{x}_\mathrm{in}-c\nabla f(\hat{x}_\mathrm{in})\right]^2\right\rangle\\ &+\gamma^2\langle \hat{p}_\mathrm{in}^2\rangle + \left\langle \Delta N^2\left(\frac{c}{\gamma},\gamma;\hat{\bm{x}}_\mathrm{a},\hat{\bm{p}}_\mathrm{a}\right)\right\rangle.\label{variance}
\end{split}
\end{equation}
As the first term does not depend on $\gamma$, $\gamma_\mathrm{opt}$ can be estimated by minimizing the sum of the second and third terms.
The third term can be calculated if the specific form of $N(\eta,\gamma;\hat{\bm{x}}_\mathrm{a},\hat{\bm{p}}_\mathrm{a})$ is known.
Once the third term is calculated, the sum of the second and third terms becomes a one-variable function of $\gamma$, which should be able to be minimized. 

Let us consider the specific case of $f(x)=(x-a)^2$, and estimate the order of magnitude of the optimal parameters by using the typical magnitude of $\hat{x}_\mathrm{in}$.
We denote $\sqrt{\langle\hat{x}^2_\mathrm{in}\rangle}=\sigma$, which we regard as the typical magnitude of $\hat{x}_\mathrm{in}$.
As the function $f(x)$ is quadratic in $x$ and its leading term is $x^2$, the order of magnitude of the variation of $f(x)$ in the range of $\pm\sigma$ can be estimated by $\sigma^2$.
We then estimate the gradient of the function by dividing the typical range of the function output $\sigma^2$ by the typical range of the function input $\sigma$: $|\langle\nabla f(x)\rangle|\sim \sigma^2/\sigma=\sigma$.
We can also assume that $|x_\mathrm{min}|\sim\sigma$.
Using these estimations, we can estimate $\eta_\mathrm{opt}\gamma_\mathrm{opt}$ as
\begin{equation}
\eta_\mathrm{opt}\gamma_\mathrm{opt}\sim 1
\end{equation}
from Eq.~(\ref{product}).
Let us then calculate the variance under $\eta\gamma=1$.
Since $N(\eta,\gamma;\hat{\bm{x}}_\mathrm{a},\hat{\bm{p}}_\mathrm{a})=-2\eta\hat{x}_\mathrm{a}$ from Eq.~(\ref{eq4}), the sum of the second and third terms of Eq.~(\ref{variance}) becomes $\gamma^2\langle \hat{p}_\mathrm{in}^2\rangle+(4/\gamma^2)\langle \hat{x}_\mathrm{a}^2\rangle$.
By minimizing this in terms of $\gamma$, $\gamma_\mathrm{opt}$ can be estimated as
\begin{equation}
\gamma_\mathrm{opt}\sim\left(4\langle\hat{x}_\mathrm{a}^2\rangle/\langle\hat{p}_\mathrm{in}^2\rangle\right)^{1/4}=\sqrt{2}\sim1.
\end{equation}
In this way, the optimal parameters are estimated as $(\eta_\mathrm{opt},\gamma_\mathrm{opt})\sim(1,1)$, which is why we set the search range by $[0.1,\,10]^{\otimes 2}$ with a margin of about a factor of ten.

The above discussion is only a rough estimation of the order of magnitude, and although it works in our case, there is no guarantee that the optimal parameters exist in the search range.
For this reason, a general strategy may be to perform the CV-QAOA within the initial search range and, if there seems to be the optimal point outside the search range, expand or change the search range by observing the behavior of the parameter optimization.

\section{REPEATED EXECUTION OF THE CIRCUITS}\label{apd}
In the demonstration of the CV-QAOA with parameters updated (Fig.~\ref{results_combined}), the iterations of the circuit execution are hierarchical.
For clarity, the conditions for that hierarchical execution are summarized here.
For each pair of the parameters, we repeatedly run the circuit and sample $\hat{x}_\mathrm{out}$ 1000 times with the parameters fixed to obtain $\langle f(\hat{x}_\mathrm{out}) \rangle$.
The Bayesian optimizer suggests 100 pairs of the parameters by using the results of $\langle f(\hat{x}_\mathrm{out}) \rangle$.
Such 100-step classical optimization is repeated eleven times for the same value of $a$ (in $f(x)=(x-a)^2$).
Finally, this repeat is done for 30 different values of $a$.

The values of $a$ are randomly sampled from the Gaussian distribution of zero mean and the standard deviation equal to 1.99, which corresponds to $\sqrt{\langle\hat{x}_\mathrm{in}^2\rangle}$.
This is because we intend to mimic the following situation.
Suppose that the range of $x$ that gives the minimum of $f(x)$ ($x_\mathrm{min}$) can be roughly estimated by some conditions in the problem settings such as, for example, physical conditions or features of $f(x)$.
In this case, the distribution of the initial state can be set so that it covers the estimated range for $x_\mathrm{min}$.
In this demonstration, mimicking the situation where the range estimation is correct and the solution $a$ is in fact in the estimated range, we repeatedly execute the CV-QAOA with many different $a$ sampled from that range to statistically evaluate the performance of the algorithm.
Note that the range for $x_\mathrm{min}$ can always be matched to the range of the initial state distribution by rescaling and translating $x$, and resultantly redefining $f(x)$.
Generally, if the estimation is incorrect, one can perform the algorithm again with an effectively broader initial state. 

\section{SIMULATION CONDITION}\label{ape}
In the simulation, the optical circuit is numerically simulated by expressing the quadratures of the initial and ancillary states using Gaussian random numbers.
Based on the measured squeezing level, we set $\langle \hat{x}_\mathrm{in}^2\rangle=\langle \hat{p}_\mathrm{a}^2\rangle=10^{9.0/10}/2$ and $\langle \hat{p}_\mathrm{in}^2\rangle=\langle \hat{x}_\mathrm{a}^2\rangle=10^{-5.3/10}/2$.
The asymmetry between the squeezing and anti-squeezing implies that optical loss is taken into account.
No other imperfections are included in the simulation.

%


\begin{thebibliography}{37}%
\makeatletter
\providecommand \@ifxundefined [1]{%
 \@ifx{#1\undefined}
}%
\providecommand \@ifnum [1]{%
 \ifnum #1\expandafter \@firstoftwo
 \else \expandafter \@secondoftwo
 \fi
}%
\providecommand \@ifx [1]{%
 \ifx #1\expandafter \@firstoftwo
 \else \expandafter \@secondoftwo
 \fi
}%
\providecommand \natexlab [1]{#1}%
\providecommand \enquote  [1]{``#1''}%
\providecommand \bibnamefont  [1]{#1}%
\providecommand \bibfnamefont [1]{#1}%
\providecommand \citenamefont [1]{#1}%
\providecommand \href@noop [0]{\@secondoftwo}%
\providecommand \href [0]{\begingroup \@sanitize@url \@href}%
\providecommand \@href[1]{\@@startlink{#1}\@@href}%
\providecommand \@@href[1]{\endgroup#1\@@endlink}%
\providecommand \@sanitize@url [0]{\catcode `\\12\catcode `\$12\catcode
  `\&12\catcode `\#12\catcode `\^12\catcode `\_12\catcode `\%12\relax}%
\providecommand \@@startlink[1]{}%
\providecommand \@@endlink[0]{}%
\providecommand \url  [0]{\begingroup\@sanitize@url \@url }%
\providecommand \@url [1]{\endgroup\@href {#1}{\urlprefix }}%
\providecommand \urlprefix  [0]{URL }%
\providecommand \Eprint [0]{\href }%
\providecommand \doibase [0]{http://dx.doi.org/}%
\providecommand \selectlanguage [0]{\@gobble}%
\providecommand \bibinfo  [0]{\@secondoftwo}%
\providecommand \bibfield  [0]{\@secondoftwo}%
\providecommand \translation [1]{[#1]}%
\providecommand \BibitemOpen [0]{}%
\providecommand \bibitemStop [0]{}%
\providecommand \bibitemNoStop [0]{.\EOS\space}%
\providecommand \EOS [0]{\spacefactor3000\relax}%
\providecommand \BibitemShut  [1]{\csname bibitem#1\endcsname}%
\let\auto@bib@innerbib\@empty
\bibitem [{\citenamefont {Preskill}(2018)}]{preskill2018}%
  \BibitemOpen
  \bibfield  {author} {\bibinfo {author} {\bibfnamefont {John}\ \bibnamefont
  {Preskill}},\ }\bibfield  {title} {\enquote {\bibinfo {title} {Quantum
  {{Computing}} in the {{NISQ}} era and beyond},}\ }\href {\doibase
  10.22331/q-2018-08-06-79} {\bibfield  {journal} {\bibinfo  {journal}
  {Quantum}\ }\textbf {\bibinfo {volume} {2}},\ \bibinfo {pages} {79} (\bibinfo
  {year} {2018})},\ \Eprint {http://arxiv.org/abs/1801.00862}
  {arXiv:1801.00862} \BibitemShut {NoStop}%
\bibitem [{\citenamefont {Cerezo}\ \emph {et~al.}(2021)\citenamefont {Cerezo},
  \citenamefont {Arrasmith}, \citenamefont {Babbush}, \citenamefont {Benjamin},
  \citenamefont {Endo}, \citenamefont {Fujii}, \citenamefont {McClean},
  \citenamefont {Mitarai}, \citenamefont {Yuan}, \citenamefont {Cincio},\ and\
  \citenamefont {Coles}}]{cerezo2021}%
  \BibitemOpen
  \bibfield  {author} {\bibinfo {author} {\bibfnamefont {M.}~\bibnamefont
  {Cerezo}}, \bibinfo {author} {\bibfnamefont {Andrew}\ \bibnamefont
  {Arrasmith}}, \bibinfo {author} {\bibfnamefont {Ryan}\ \bibnamefont
  {Babbush}}, \bibinfo {author} {\bibfnamefont {Simon~C.}\ \bibnamefont
  {Benjamin}}, \bibinfo {author} {\bibfnamefont {Suguru}\ \bibnamefont {Endo}},
  \bibinfo {author} {\bibfnamefont {Keisuke}\ \bibnamefont {Fujii}}, \bibinfo
  {author} {\bibfnamefont {Jarrod~R.}\ \bibnamefont {McClean}}, \bibinfo
  {author} {\bibfnamefont {Kosuke}\ \bibnamefont {Mitarai}}, \bibinfo {author}
  {\bibfnamefont {Xiao}\ \bibnamefont {Yuan}}, \bibinfo {author} {\bibfnamefont
  {Lukasz}\ \bibnamefont {Cincio}}, \ and\ \bibinfo {author} {\bibfnamefont
  {Patrick~J.}\ \bibnamefont {Coles}},\ }\bibfield  {title} {\enquote {\bibinfo
  {title} {Variational quantum algorithms},}\ }\href {\doibase
  10.1038/s42254-021-00348-9} {\bibfield  {journal} {\bibinfo  {journal}
  {Nature Reviews Physics}\ }\textbf {\bibinfo {volume} {3}},\ \bibinfo {pages}
  {625--644} (\bibinfo {year} {2021})},\ \Eprint
  {http://arxiv.org/abs/2012.09265} {arXiv:2012.09265} \BibitemShut {NoStop}%
\bibitem [{\citenamefont {Farhi}\ \emph {et~al.}(2014)\citenamefont {Farhi},
  \citenamefont {Goldstone},\ and\ \citenamefont {Gutmann}}]{farhi2014}%
  \BibitemOpen
  \bibfield  {author} {\bibinfo {author} {\bibfnamefont {Edward}\ \bibnamefont
  {Farhi}}, \bibinfo {author} {\bibfnamefont {Jeffrey}\ \bibnamefont
  {Goldstone}}, \ and\ \bibinfo {author} {\bibfnamefont {Sam}\ \bibnamefont
  {Gutmann}},\ }\bibfield  {title} {\enquote {\bibinfo {title} {A {{Quantum
  Approximate Optimization Algorithm}}},}\ }\href@noop {} {\bibfield  {journal}
  {\bibinfo  {journal} {arXiv:1411.4028 [quant-ph]}\ } (\bibinfo {year}
  {2014})},\ \Eprint {http://arxiv.org/abs/1411.4028} {arXiv:1411.4028
  [quant-ph]} \BibitemShut {NoStop}%
\bibitem [{\citenamefont {Peruzzo}\ \emph {et~al.}(2014)\citenamefont
  {Peruzzo}, \citenamefont {McClean}, \citenamefont {Shadbolt}, \citenamefont
  {Yung}, \citenamefont {Zhou}, \citenamefont {Love}, \citenamefont
  {{Aspuru-Guzik}},\ and\ \citenamefont {O'Brien}}]{peruzzo2014}%
  \BibitemOpen
  \bibfield  {author} {\bibinfo {author} {\bibfnamefont {Alberto}\ \bibnamefont
  {Peruzzo}}, \bibinfo {author} {\bibfnamefont {Jarrod}\ \bibnamefont
  {McClean}}, \bibinfo {author} {\bibfnamefont {Peter}\ \bibnamefont
  {Shadbolt}}, \bibinfo {author} {\bibfnamefont {Man-Hong}\ \bibnamefont
  {Yung}}, \bibinfo {author} {\bibfnamefont {Xiao-Qi}\ \bibnamefont {Zhou}},
  \bibinfo {author} {\bibfnamefont {Peter~J.}\ \bibnamefont {Love}}, \bibinfo
  {author} {\bibfnamefont {Al{\'a}n}\ \bibnamefont {{Aspuru-Guzik}}}, \ and\
  \bibinfo {author} {\bibfnamefont {Jeremy~L.}\ \bibnamefont {O'Brien}},\
  }\bibfield  {title} {\enquote {\bibinfo {title} {A variational eigenvalue
  solver on a photonic quantum processor},}\ }\href {\doibase
  10.1038/ncomms5213} {\bibfield  {journal} {\bibinfo  {journal} {Nature
  Communications}\ }\textbf {\bibinfo {volume} {5}},\ \bibinfo {pages} {4213}
  (\bibinfo {year} {2014})}\BibitemShut {NoStop}%
\bibitem [{\citenamefont {Mitarai}\ \emph {et~al.}(2018)\citenamefont
  {Mitarai}, \citenamefont {Negoro}, \citenamefont {Kitagawa},\ and\
  \citenamefont {Fujii}}]{mitarai2018}%
  \BibitemOpen
  \bibfield  {author} {\bibinfo {author} {\bibfnamefont {K.}~\bibnamefont
  {Mitarai}}, \bibinfo {author} {\bibfnamefont {M.}~\bibnamefont {Negoro}},
  \bibinfo {author} {\bibfnamefont {M.}~\bibnamefont {Kitagawa}}, \ and\
  \bibinfo {author} {\bibfnamefont {K.}~\bibnamefont {Fujii}},\ }\bibfield
  {title} {\enquote {\bibinfo {title} {Quantum circuit learning},}\ }\href
  {\doibase 10.1103/PhysRevA.98.032309} {\bibfield  {journal} {\bibinfo
  {journal} {Physical Review A}\ }\textbf {\bibinfo {volume} {98}},\ \bibinfo
  {pages} {032309} (\bibinfo {year} {2018})}\BibitemShut {NoStop}%
\bibitem [{\citenamefont {Otterbach}\ \emph {et~al.}(2017)\citenamefont
  {Otterbach}, \citenamefont {Manenti}, \citenamefont {Alidoust}, \citenamefont
  {Bestwick}, \citenamefont {Block}, \citenamefont {Bloom}, \citenamefont
  {Caldwell}, \citenamefont {Didier}, \citenamefont {Fried}, \citenamefont
  {Hong}, \citenamefont {Karalekas}, \citenamefont {Osborn}, \citenamefont
  {Papageorge}, \citenamefont {Peterson}, \citenamefont {Prawiroatmodjo},
  \citenamefont {Rubin}, \citenamefont {Ryan}, \citenamefont {Scarabelli},
  \citenamefont {Scheer}, \citenamefont {Sete}, \citenamefont {Sivarajah},
  \citenamefont {Smith}, \citenamefont {Staley}, \citenamefont {Tezak},
  \citenamefont {Zeng}, \citenamefont {Hudson}, \citenamefont {Johnson},
  \citenamefont {Reagor}, \citenamefont {{da Silva}},\ and\ \citenamefont
  {Rigetti}}]{otterbach2017}%
  \BibitemOpen
  \bibfield  {author} {\bibinfo {author} {\bibfnamefont {J.~S.}\ \bibnamefont
  {Otterbach}}, \bibinfo {author} {\bibfnamefont {R.}~\bibnamefont {Manenti}},
  \bibinfo {author} {\bibfnamefont {N.}~\bibnamefont {Alidoust}}, \bibinfo
  {author} {\bibfnamefont {A.}~\bibnamefont {Bestwick}}, \bibinfo {author}
  {\bibfnamefont {M.}~\bibnamefont {Block}}, \bibinfo {author} {\bibfnamefont
  {B.}~\bibnamefont {Bloom}}, \bibinfo {author} {\bibfnamefont
  {S.}~\bibnamefont {Caldwell}}, \bibinfo {author} {\bibfnamefont
  {N.}~\bibnamefont {Didier}}, \bibinfo {author} {\bibfnamefont {E.~Schuyler}\
  \bibnamefont {Fried}}, \bibinfo {author} {\bibfnamefont {S.}~\bibnamefont
  {Hong}}, \bibinfo {author} {\bibfnamefont {P.}~\bibnamefont {Karalekas}},
  \bibinfo {author} {\bibfnamefont {C.~B.}\ \bibnamefont {Osborn}}, \bibinfo
  {author} {\bibfnamefont {A.}~\bibnamefont {Papageorge}}, \bibinfo {author}
  {\bibfnamefont {E.~C.}\ \bibnamefont {Peterson}}, \bibinfo {author}
  {\bibfnamefont {G.}~\bibnamefont {Prawiroatmodjo}}, \bibinfo {author}
  {\bibfnamefont {N.}~\bibnamefont {Rubin}}, \bibinfo {author} {\bibfnamefont
  {Colm~A.}\ \bibnamefont {Ryan}}, \bibinfo {author} {\bibfnamefont
  {D.}~\bibnamefont {Scarabelli}}, \bibinfo {author} {\bibfnamefont
  {M.}~\bibnamefont {Scheer}}, \bibinfo {author} {\bibfnamefont {E.~A.}\
  \bibnamefont {Sete}}, \bibinfo {author} {\bibfnamefont {P.}~\bibnamefont
  {Sivarajah}}, \bibinfo {author} {\bibfnamefont {Robert~S.}\ \bibnamefont
  {Smith}}, \bibinfo {author} {\bibfnamefont {A.}~\bibnamefont {Staley}},
  \bibinfo {author} {\bibfnamefont {N.}~\bibnamefont {Tezak}}, \bibinfo
  {author} {\bibfnamefont {W.~J.}\ \bibnamefont {Zeng}}, \bibinfo {author}
  {\bibfnamefont {A.}~\bibnamefont {Hudson}}, \bibinfo {author} {\bibfnamefont
  {Blake~R.}\ \bibnamefont {Johnson}}, \bibinfo {author} {\bibfnamefont
  {M.}~\bibnamefont {Reagor}}, \bibinfo {author} {\bibfnamefont {M.~P.}\
  \bibnamefont {{da Silva}}}, \ and\ \bibinfo {author} {\bibfnamefont
  {C.}~\bibnamefont {Rigetti}},\ }\bibfield  {title} {\enquote {\bibinfo
  {title} {Unsupervised {{Machine Learning}} on a {{Hybrid Quantum
  Computer}}},}\ }\href@noop {} {\bibfield  {journal} {\bibinfo  {journal}
  {arXiv:1712.05771 [quant-ph]}\ } (\bibinfo {year} {2017})},\ \Eprint
  {http://arxiv.org/abs/1712.05771} {arXiv:1712.05771 [quant-ph]} \BibitemShut
  {NoStop}%
\bibitem [{\citenamefont {Bengtsson}\ \emph {et~al.}(2020)\citenamefont
  {Bengtsson}, \citenamefont {Vikst{\aa}l}, \citenamefont {Warren},
  \citenamefont {Svensson}, \citenamefont {Gu}, \citenamefont {Kockum},
  \citenamefont {Krantz}, \citenamefont {Kri{\v z}an}, \citenamefont {Shiri},
  \citenamefont {Svensson}, \citenamefont {Tancredi}, \citenamefont
  {Johansson}, \citenamefont {Delsing}, \citenamefont {Ferrini},\ and\
  \citenamefont {Bylander}}]{bengtsson2020}%
  \BibitemOpen
  \bibfield  {author} {\bibinfo {author} {\bibfnamefont {Andreas}\ \bibnamefont
  {Bengtsson}}, \bibinfo {author} {\bibfnamefont {Pontus}\ \bibnamefont
  {Vikst{\aa}l}}, \bibinfo {author} {\bibfnamefont {Christopher}\ \bibnamefont
  {Warren}}, \bibinfo {author} {\bibfnamefont {Marika}\ \bibnamefont
  {Svensson}}, \bibinfo {author} {\bibfnamefont {Xiu}\ \bibnamefont {Gu}},
  \bibinfo {author} {\bibfnamefont {Anton~Frisk}\ \bibnamefont {Kockum}},
  \bibinfo {author} {\bibfnamefont {Philip}\ \bibnamefont {Krantz}}, \bibinfo
  {author} {\bibfnamefont {Christian}\ \bibnamefont {Kri{\v z}an}}, \bibinfo
  {author} {\bibfnamefont {Daryoush}\ \bibnamefont {Shiri}}, \bibinfo {author}
  {\bibfnamefont {Ida-Maria}\ \bibnamefont {Svensson}}, \bibinfo {author}
  {\bibfnamefont {Giovanna}\ \bibnamefont {Tancredi}}, \bibinfo {author}
  {\bibfnamefont {G{\"o}ran}\ \bibnamefont {Johansson}}, \bibinfo {author}
  {\bibfnamefont {Per}\ \bibnamefont {Delsing}}, \bibinfo {author}
  {\bibfnamefont {Giulia}\ \bibnamefont {Ferrini}}, \ and\ \bibinfo {author}
  {\bibfnamefont {Jonas}\ \bibnamefont {Bylander}},\ }\bibfield  {title}
  {\enquote {\bibinfo {title} {Improved {{Success Probability}} with {{Greater
  Circuit Depth}} for the {{Quantum Approximate Optimization Algorithm}}},}\
  }\href {\doibase 10.1103/PhysRevApplied.14.034010} {\bibfield  {journal}
  {\bibinfo  {journal} {Physical Review Applied}\ }\textbf {\bibinfo {volume}
  {14}},\ \bibinfo {pages} {034010} (\bibinfo {year} {2020})}\BibitemShut
  {NoStop}%
\bibitem [{\citenamefont {Pagano}\ \emph {et~al.}(2020)\citenamefont {Pagano},
  \citenamefont {Bapat}, \citenamefont {Becker}, \citenamefont {Collins},
  \citenamefont {De}, \citenamefont {Hess}, \citenamefont {Kaplan},
  \citenamefont {Kyprianidis}, \citenamefont {Tan}, \citenamefont {Baldwin},
  \citenamefont {Brady}, \citenamefont {Deshpande}, \citenamefont {Liu},
  \citenamefont {Jordan}, \citenamefont {Gorshkov},\ and\ \citenamefont
  {Monroe}}]{pagano2020}%
  \BibitemOpen
  \bibfield  {author} {\bibinfo {author} {\bibfnamefont {Guido}\ \bibnamefont
  {Pagano}}, \bibinfo {author} {\bibfnamefont {Aniruddha}\ \bibnamefont
  {Bapat}}, \bibinfo {author} {\bibfnamefont {Patrick}\ \bibnamefont {Becker}},
  \bibinfo {author} {\bibfnamefont {Katherine~S.}\ \bibnamefont {Collins}},
  \bibinfo {author} {\bibfnamefont {Arinjoy}\ \bibnamefont {De}}, \bibinfo
  {author} {\bibfnamefont {Paul~W.}\ \bibnamefont {Hess}}, \bibinfo {author}
  {\bibfnamefont {Harvey~B.}\ \bibnamefont {Kaplan}}, \bibinfo {author}
  {\bibfnamefont {Antonis}\ \bibnamefont {Kyprianidis}}, \bibinfo {author}
  {\bibfnamefont {Wen~Lin}\ \bibnamefont {Tan}}, \bibinfo {author}
  {\bibfnamefont {Christopher}\ \bibnamefont {Baldwin}}, \bibinfo {author}
  {\bibfnamefont {Lucas~T.}\ \bibnamefont {Brady}}, \bibinfo {author}
  {\bibfnamefont {Abhinav}\ \bibnamefont {Deshpande}}, \bibinfo {author}
  {\bibfnamefont {Fangli}\ \bibnamefont {Liu}}, \bibinfo {author}
  {\bibfnamefont {Stephen}\ \bibnamefont {Jordan}}, \bibinfo {author}
  {\bibfnamefont {Alexey~V.}\ \bibnamefont {Gorshkov}}, \ and\ \bibinfo
  {author} {\bibfnamefont {Christopher}\ \bibnamefont {Monroe}},\ }\bibfield
  {title} {\enquote {\bibinfo {title} {Quantum approximate optimization of the
  long-range {{Ising}} model with a trapped-ion quantum simulator},}\ }\href
  {\doibase 10.1073/pnas.2006373117} {\bibfield  {journal} {\bibinfo  {journal}
  {Proceedings of the National Academy of Sciences}\ }\textbf {\bibinfo
  {volume} {117}},\ \bibinfo {pages} {25396--25401} (\bibinfo {year}
  {2020})}\BibitemShut {NoStop}%
\bibitem [{\citenamefont {Harrigan}\ \emph {et~al.}(2021)\citenamefont
  {Harrigan}, \citenamefont {Sung}, \citenamefont {Neeley}, \citenamefont
  {Satzinger}, \citenamefont {Arute}, \citenamefont {Arya}, \citenamefont
  {Atalaya}, \citenamefont {Bardin}, \citenamefont {Barends}, \citenamefont
  {Boixo}, \citenamefont {Broughton}, \citenamefont {Buckley}, \citenamefont
  {Buell}, \citenamefont {Burkett}, \citenamefont {Bushnell}, \citenamefont
  {Chen}, \citenamefont {Chen}, \citenamefont {{Ben Chiaro}}, \citenamefont
  {Collins}, \citenamefont {Courtney}, \citenamefont {Demura}, \citenamefont
  {Dunsworth}, \citenamefont {Eppens}, \citenamefont {Fowler}, \citenamefont
  {Foxen}, \citenamefont {Gidney}, \citenamefont {Giustina}, \citenamefont
  {Graff}, \citenamefont {Habegger}, \citenamefont {Ho}, \citenamefont {Hong},
  \citenamefont {Huang}, \citenamefont {Ioffe}, \citenamefont {Isakov},
  \citenamefont {Jeffrey}, \citenamefont {Jiang}, \citenamefont {Jones},
  \citenamefont {Kafri}, \citenamefont {Kechedzhi}, \citenamefont {Kelly},
  \citenamefont {Kim}, \citenamefont {Klimov}, \citenamefont {Korotkov},
  \citenamefont {Kostritsa}, \citenamefont {Landhuis}, \citenamefont {Laptev},
  \citenamefont {Lindmark}, \citenamefont {Leib}, \citenamefont {Martin},
  \citenamefont {Martinis}, \citenamefont {McClean}, \citenamefont {McEwen},
  \citenamefont {Megrant}, \citenamefont {Mi}, \citenamefont {Mohseni},
  \citenamefont {Mruczkiewicz}, \citenamefont {Mutus}, \citenamefont {Naaman},
  \citenamefont {Neill}, \citenamefont {Neukart}, \citenamefont {Niu},
  \citenamefont {O'Brien}, \citenamefont {O'Gorman}, \citenamefont {Ostby},
  \citenamefont {Petukhov}, \citenamefont {Putterman}, \citenamefont
  {Quintana}, \citenamefont {Roushan}, \citenamefont {Rubin}, \citenamefont
  {Sank}, \citenamefont {Skolik}, \citenamefont {Smelyanskiy}, \citenamefont
  {Strain}, \citenamefont {Streif}, \citenamefont {Szalay}, \citenamefont
  {Vainsencher}, \citenamefont {White}, \citenamefont {Yao}, \citenamefont
  {Yeh}, \citenamefont {Zalcman}, \citenamefont {Zhou}, \citenamefont {Neven},
  \citenamefont {Bacon}, \citenamefont {Lucero}, \citenamefont {Farhi},\ and\
  \citenamefont {Babbush}}]{harrigan2021}%
  \BibitemOpen
  \bibfield  {author} {\bibinfo {author} {\bibfnamefont {Matthew~P.}\
  \bibnamefont {Harrigan}}, \bibinfo {author} {\bibfnamefont {Kevin~J.}\
  \bibnamefont {Sung}}, \bibinfo {author} {\bibfnamefont {Matthew}\
  \bibnamefont {Neeley}}, \bibinfo {author} {\bibfnamefont {Kevin~J.}\
  \bibnamefont {Satzinger}}, \bibinfo {author} {\bibfnamefont {Frank}\
  \bibnamefont {Arute}}, \bibinfo {author} {\bibfnamefont {Kunal}\ \bibnamefont
  {Arya}}, \bibinfo {author} {\bibfnamefont {Juan}\ \bibnamefont {Atalaya}},
  \bibinfo {author} {\bibfnamefont {Joseph~C.}\ \bibnamefont {Bardin}},
  \bibinfo {author} {\bibfnamefont {Rami}\ \bibnamefont {Barends}}, \bibinfo
  {author} {\bibfnamefont {Sergio}\ \bibnamefont {Boixo}}, \bibinfo {author}
  {\bibfnamefont {Michael}\ \bibnamefont {Broughton}}, \bibinfo {author}
  {\bibfnamefont {Bob~B.}\ \bibnamefont {Buckley}}, \bibinfo {author}
  {\bibfnamefont {David~A.}\ \bibnamefont {Buell}}, \bibinfo {author}
  {\bibfnamefont {Brian}\ \bibnamefont {Burkett}}, \bibinfo {author}
  {\bibfnamefont {Nicholas}\ \bibnamefont {Bushnell}}, \bibinfo {author}
  {\bibfnamefont {Yu}~\bibnamefont {Chen}}, \bibinfo {author} {\bibfnamefont
  {Zijun}\ \bibnamefont {Chen}}, \bibinfo {author} {\bibnamefont {{Ben
  Chiaro}}}, \bibinfo {author} {\bibfnamefont {Roberto}\ \bibnamefont
  {Collins}}, \bibinfo {author} {\bibfnamefont {William}\ \bibnamefont
  {Courtney}}, \bibinfo {author} {\bibfnamefont {Sean}\ \bibnamefont {Demura}},
  \bibinfo {author} {\bibfnamefont {Andrew}\ \bibnamefont {Dunsworth}},
  \bibinfo {author} {\bibfnamefont {Daniel}\ \bibnamefont {Eppens}}, \bibinfo
  {author} {\bibfnamefont {Austin}\ \bibnamefont {Fowler}}, \bibinfo {author}
  {\bibfnamefont {Brooks}\ \bibnamefont {Foxen}}, \bibinfo {author}
  {\bibfnamefont {Craig}\ \bibnamefont {Gidney}}, \bibinfo {author}
  {\bibfnamefont {Marissa}\ \bibnamefont {Giustina}}, \bibinfo {author}
  {\bibfnamefont {Rob}\ \bibnamefont {Graff}}, \bibinfo {author} {\bibfnamefont
  {Steve}\ \bibnamefont {Habegger}}, \bibinfo {author} {\bibfnamefont {Alan}\
  \bibnamefont {Ho}}, \bibinfo {author} {\bibfnamefont {Sabrina}\ \bibnamefont
  {Hong}}, \bibinfo {author} {\bibfnamefont {Trent}\ \bibnamefont {Huang}},
  \bibinfo {author} {\bibfnamefont {L.~B.}\ \bibnamefont {Ioffe}}, \bibinfo
  {author} {\bibfnamefont {Sergei~V.}\ \bibnamefont {Isakov}}, \bibinfo
  {author} {\bibfnamefont {Evan}\ \bibnamefont {Jeffrey}}, \bibinfo {author}
  {\bibfnamefont {Zhang}\ \bibnamefont {Jiang}}, \bibinfo {author}
  {\bibfnamefont {Cody}\ \bibnamefont {Jones}}, \bibinfo {author}
  {\bibfnamefont {Dvir}\ \bibnamefont {Kafri}}, \bibinfo {author}
  {\bibfnamefont {Kostyantyn}\ \bibnamefont {Kechedzhi}}, \bibinfo {author}
  {\bibfnamefont {Julian}\ \bibnamefont {Kelly}}, \bibinfo {author}
  {\bibfnamefont {Seon}\ \bibnamefont {Kim}}, \bibinfo {author} {\bibfnamefont
  {Paul~V.}\ \bibnamefont {Klimov}}, \bibinfo {author} {\bibfnamefont
  {Alexander~N.}\ \bibnamefont {Korotkov}}, \bibinfo {author} {\bibfnamefont
  {Fedor}\ \bibnamefont {Kostritsa}}, \bibinfo {author} {\bibfnamefont {David}\
  \bibnamefont {Landhuis}}, \bibinfo {author} {\bibfnamefont {Pavel}\
  \bibnamefont {Laptev}}, \bibinfo {author} {\bibfnamefont {Mike}\ \bibnamefont
  {Lindmark}}, \bibinfo {author} {\bibfnamefont {Martin}\ \bibnamefont {Leib}},
  \bibinfo {author} {\bibfnamefont {Orion}\ \bibnamefont {Martin}}, \bibinfo
  {author} {\bibfnamefont {John~M.}\ \bibnamefont {Martinis}}, \bibinfo
  {author} {\bibfnamefont {Jarrod~R.}\ \bibnamefont {McClean}}, \bibinfo
  {author} {\bibfnamefont {Matt}\ \bibnamefont {McEwen}}, \bibinfo {author}
  {\bibfnamefont {Anthony}\ \bibnamefont {Megrant}}, \bibinfo {author}
  {\bibfnamefont {Xiao}\ \bibnamefont {Mi}}, \bibinfo {author} {\bibfnamefont
  {Masoud}\ \bibnamefont {Mohseni}}, \bibinfo {author} {\bibfnamefont
  {Wojciech}\ \bibnamefont {Mruczkiewicz}}, \bibinfo {author} {\bibfnamefont
  {Josh}\ \bibnamefont {Mutus}}, \bibinfo {author} {\bibfnamefont {Ofer}\
  \bibnamefont {Naaman}}, \bibinfo {author} {\bibfnamefont {Charles}\
  \bibnamefont {Neill}}, \bibinfo {author} {\bibfnamefont {Florian}\
  \bibnamefont {Neukart}}, \bibinfo {author} {\bibfnamefont {Murphy~Yuezhen}\
  \bibnamefont {Niu}}, \bibinfo {author} {\bibfnamefont {Thomas~E.}\
  \bibnamefont {O'Brien}}, \bibinfo {author} {\bibfnamefont {Bryan}\
  \bibnamefont {O'Gorman}}, \bibinfo {author} {\bibfnamefont {Eric}\
  \bibnamefont {Ostby}}, \bibinfo {author} {\bibfnamefont {Andre}\ \bibnamefont
  {Petukhov}}, \bibinfo {author} {\bibfnamefont {Harald}\ \bibnamefont
  {Putterman}}, \bibinfo {author} {\bibfnamefont {Chris}\ \bibnamefont
  {Quintana}}, \bibinfo {author} {\bibfnamefont {Pedram}\ \bibnamefont
  {Roushan}}, \bibinfo {author} {\bibfnamefont {Nicholas~C.}\ \bibnamefont
  {Rubin}}, \bibinfo {author} {\bibfnamefont {Daniel}\ \bibnamefont {Sank}},
  \bibinfo {author} {\bibfnamefont {Andrea}\ \bibnamefont {Skolik}}, \bibinfo
  {author} {\bibfnamefont {Vadim}\ \bibnamefont {Smelyanskiy}}, \bibinfo
  {author} {\bibfnamefont {Doug}\ \bibnamefont {Strain}}, \bibinfo {author}
  {\bibfnamefont {Michael}\ \bibnamefont {Streif}}, \bibinfo {author}
  {\bibfnamefont {Marco}\ \bibnamefont {Szalay}}, \bibinfo {author}
  {\bibfnamefont {Amit}\ \bibnamefont {Vainsencher}}, \bibinfo {author}
  {\bibfnamefont {Theodore}\ \bibnamefont {White}}, \bibinfo {author}
  {\bibfnamefont {Z.~Jamie}\ \bibnamefont {Yao}}, \bibinfo {author}
  {\bibfnamefont {Ping}\ \bibnamefont {Yeh}}, \bibinfo {author} {\bibfnamefont
  {Adam}\ \bibnamefont {Zalcman}}, \bibinfo {author} {\bibfnamefont {Leo}\
  \bibnamefont {Zhou}}, \bibinfo {author} {\bibfnamefont {Hartmut}\
  \bibnamefont {Neven}}, \bibinfo {author} {\bibfnamefont {Dave}\ \bibnamefont
  {Bacon}}, \bibinfo {author} {\bibfnamefont {Erik}\ \bibnamefont {Lucero}},
  \bibinfo {author} {\bibfnamefont {Edward}\ \bibnamefont {Farhi}}, \ and\
  \bibinfo {author} {\bibfnamefont {Ryan}\ \bibnamefont {Babbush}},\ }\bibfield
   {title} {\enquote {\bibinfo {title} {Quantum approximate optimization of
  non-planar graph problems on a planar superconducting processor},}\ }\href
  {\doibase 10.1038/s41567-020-01105-y} {\bibfield  {journal} {\bibinfo
  {journal} {Nature Physics}\ }\textbf {\bibinfo {volume} {17}},\ \bibinfo
  {pages} {332--336} (\bibinfo {year} {2021})}\BibitemShut {NoStop}%
\bibitem [{\citenamefont {Havl{\'i}{\v c}ek}\ \emph {et~al.}(2019)\citenamefont
  {Havl{\'i}{\v c}ek}, \citenamefont {C{\'o}rcoles}, \citenamefont {Temme},
  \citenamefont {Harrow}, \citenamefont {Kandala}, \citenamefont {Chow},\ and\
  \citenamefont {Gambetta}}]{havlicek2019}%
  \BibitemOpen
  \bibfield  {author} {\bibinfo {author} {\bibfnamefont {Vojt{\v e}ch}\
  \bibnamefont {Havl{\'i}{\v c}ek}}, \bibinfo {author} {\bibfnamefont
  {Antonio~D.}\ \bibnamefont {C{\'o}rcoles}}, \bibinfo {author} {\bibfnamefont
  {Kristan}\ \bibnamefont {Temme}}, \bibinfo {author} {\bibfnamefont {Aram~W.}\
  \bibnamefont {Harrow}}, \bibinfo {author} {\bibfnamefont {Abhinav}\
  \bibnamefont {Kandala}}, \bibinfo {author} {\bibfnamefont {Jerry~M.}\
  \bibnamefont {Chow}}, \ and\ \bibinfo {author} {\bibfnamefont {Jay~M.}\
  \bibnamefont {Gambetta}},\ }\bibfield  {title} {\enquote {\bibinfo {title}
  {Supervised learning with quantum-enhanced feature spaces},}\ }\href
  {\doibase 10.1038/s41586-019-0980-2} {\bibfield  {journal} {\bibinfo
  {journal} {Nature}\ }\textbf {\bibinfo {volume} {567}},\ \bibinfo {pages}
  {209--212} (\bibinfo {year} {2019})}\BibitemShut {NoStop}%
\bibitem [{\citenamefont {Verdon}\ \emph {et~al.}(2019)\citenamefont {Verdon},
  \citenamefont {Arrazola}, \citenamefont {Br{\'a}dler},\ and\ \citenamefont
  {Killoran}}]{verdon2019}%
  \BibitemOpen
  \bibfield  {author} {\bibinfo {author} {\bibfnamefont {Guillaume}\
  \bibnamefont {Verdon}}, \bibinfo {author} {\bibfnamefont {Juan~Miguel}\
  \bibnamefont {Arrazola}}, \bibinfo {author} {\bibfnamefont {Kamil}\
  \bibnamefont {Br{\'a}dler}}, \ and\ \bibinfo {author} {\bibfnamefont
  {Nathan}\ \bibnamefont {Killoran}},\ }\bibfield  {title} {\enquote {\bibinfo
  {title} {A {{Quantum Approximate Optimization Algorithm}} for continuous
  problems},}\ }\href@noop {} {\bibfield  {journal} {\bibinfo  {journal}
  {arXiv:1902.00409 [quant-ph]}\ } (\bibinfo {year} {2019})},\ \Eprint
  {http://arxiv.org/abs/1902.00409} {arXiv:1902.00409 [quant-ph]} \BibitemShut
  {NoStop}%
\bibitem [{\citenamefont {Killoran}\ \emph {et~al.}(2019)\citenamefont
  {Killoran}, \citenamefont {Bromley}, \citenamefont {Arrazola}, \citenamefont
  {Schuld}, \citenamefont {Quesada},\ and\ \citenamefont
  {Lloyd}}]{killoran2019}%
  \BibitemOpen
  \bibfield  {author} {\bibinfo {author} {\bibfnamefont {Nathan}\ \bibnamefont
  {Killoran}}, \bibinfo {author} {\bibfnamefont {Thomas~R.}\ \bibnamefont
  {Bromley}}, \bibinfo {author} {\bibfnamefont {Juan~Miguel}\ \bibnamefont
  {Arrazola}}, \bibinfo {author} {\bibfnamefont {Maria}\ \bibnamefont
  {Schuld}}, \bibinfo {author} {\bibfnamefont {Nicol{\'a}s}\ \bibnamefont
  {Quesada}}, \ and\ \bibinfo {author} {\bibfnamefont {Seth}\ \bibnamefont
  {Lloyd}},\ }\bibfield  {title} {\enquote {\bibinfo {title}
  {Continuous-variable quantum neural networks},}\ }\href {\doibase
  10.1103/PhysRevResearch.1.033063} {\bibfield  {journal} {\bibinfo  {journal}
  {Physical Review Research}\ }\textbf {\bibinfo {volume} {1}},\ \bibinfo
  {pages} {033063} (\bibinfo {year} {2019})}\BibitemShut {NoStop}%
\bibitem [{\citenamefont {Arrazola}\ \emph {et~al.}(2019)\citenamefont
  {Arrazola}, \citenamefont {Bromley}, \citenamefont {Izaac}, \citenamefont
  {Myers}, \citenamefont {Br{\'a}dler},\ and\ \citenamefont
  {Killoran}}]{arrazola2019}%
  \BibitemOpen
  \bibfield  {author} {\bibinfo {author} {\bibfnamefont {Juan~Miguel}\
  \bibnamefont {Arrazola}}, \bibinfo {author} {\bibfnamefont {Thomas~R.}\
  \bibnamefont {Bromley}}, \bibinfo {author} {\bibfnamefont {Josh}\
  \bibnamefont {Izaac}}, \bibinfo {author} {\bibfnamefont {Casey~R.}\
  \bibnamefont {Myers}}, \bibinfo {author} {\bibfnamefont {Kamil}\ \bibnamefont
  {Br{\'a}dler}}, \ and\ \bibinfo {author} {\bibfnamefont {Nathan}\
  \bibnamefont {Killoran}},\ }\bibfield  {title} {\enquote {\bibinfo {title}
  {Machine learning method for state preparation and gate synthesis on photonic
  quantum computers},}\ }\href {\doibase 10.1088/2058-9565/aaf59e} {\bibfield
  {journal} {\bibinfo  {journal} {Quantum Science and Technology}\ }\textbf
  {\bibinfo {volume} {4}},\ \bibinfo {pages} {024004} (\bibinfo {year}
  {2019})}\BibitemShut {NoStop}%
\bibitem [{\citenamefont {Volkoff}\ \emph {et~al.}(2021)\citenamefont
  {Volkoff}, \citenamefont {Holmes},\ and\ \citenamefont
  {Sornborger}}]{volkoff2021}%
  \BibitemOpen
  \bibfield  {author} {\bibinfo {author} {\bibfnamefont {Tyler}\ \bibnamefont
  {Volkoff}}, \bibinfo {author} {\bibfnamefont {Zo{\"e}}\ \bibnamefont
  {Holmes}}, \ and\ \bibinfo {author} {\bibfnamefont {Andrew}\ \bibnamefont
  {Sornborger}},\ }\bibfield  {title} {\enquote {\bibinfo {title} {Universal
  {{Compiling}} and ({{No-}}){{Free-Lunch Theorems}} for {{Continuous-Variable
  Quantum Learning}}},}\ }\href {\doibase 10.1103/PRXQuantum.2.040327}
  {\bibfield  {journal} {\bibinfo  {journal} {PRX Quantum}\ }\textbf {\bibinfo
  {volume} {2}},\ \bibinfo {pages} {040327} (\bibinfo {year}
  {2021})}\BibitemShut {NoStop}%
\bibitem [{\citenamefont {St{\k{e}}ch{\l}y}\ \emph {et~al.}(2019)\citenamefont
  {St{\k{e}}ch{\l}y}, \citenamefont {Bashige},\ and\ \citenamefont
  {Chojecki}}]{stechly2019}%
  \BibitemOpen
  \bibfield  {author} {\bibinfo {author} {\bibfnamefont {Micha{\l}}\
  \bibnamefont {St{\k{e}}ch{\l}y}}, \bibinfo {author} {\bibfnamefont {Ntwali}\
  \bibnamefont {Bashige}}, \ and\ \bibinfo {author} {\bibfnamefont
  {Przemys{\l}aw}\ \bibnamefont {Chojecki}},\ }\bibfield  {title} {\enquote
  {\bibinfo {title} {Approaching graph problems with continuous variable
  quantum computing},}\ }\href@noop {} {\bibfield  {journal} {\bibinfo
  {journal} {arXiv:1906.07047 [quant-ph]}\ } (\bibinfo {year} {2019})},\
  \Eprint {http://arxiv.org/abs/1906.07047} {arXiv:1906.07047 [quant-ph]}
  \BibitemShut {NoStop}%
\bibitem [{\citenamefont {Takeda}\ and\ \citenamefont
  {Furusawa}(2019)}]{takeda2019a}%
  \BibitemOpen
  \bibfield  {author} {\bibinfo {author} {\bibfnamefont {S.}~\bibnamefont
  {Takeda}}\ and\ \bibinfo {author} {\bibfnamefont {A.}~\bibnamefont
  {Furusawa}},\ }\bibfield  {title} {\enquote {\bibinfo {title} {Toward
  large-scale fault-tolerant universal photonic quantum computing},}\ }\href
  {\doibase 10.1063/1.5100160} {\bibfield  {journal} {\bibinfo  {journal} {APL
  Photonics}\ }\textbf {\bibinfo {volume} {4}},\ \bibinfo {pages} {060902}
  (\bibinfo {year} {2019})}\BibitemShut {NoStop}%
\bibitem [{\citenamefont {Hamilton}\ \emph {et~al.}(2017)\citenamefont
  {Hamilton}, \citenamefont {Kruse}, \citenamefont {Sansoni}, \citenamefont
  {Barkhofen}, \citenamefont {Silberhorn},\ and\ \citenamefont
  {Jex}}]{hamilton2017}%
  \BibitemOpen
  \bibfield  {author} {\bibinfo {author} {\bibfnamefont {Craig~S.}\
  \bibnamefont {Hamilton}}, \bibinfo {author} {\bibfnamefont {Regina}\
  \bibnamefont {Kruse}}, \bibinfo {author} {\bibfnamefont {Linda}\ \bibnamefont
  {Sansoni}}, \bibinfo {author} {\bibfnamefont {Sonja}\ \bibnamefont
  {Barkhofen}}, \bibinfo {author} {\bibfnamefont {Christine}\ \bibnamefont
  {Silberhorn}}, \ and\ \bibinfo {author} {\bibfnamefont {Igor}\ \bibnamefont
  {Jex}},\ }\bibfield  {title} {\enquote {\bibinfo {title} {Gaussian {{Boson
  Sampling}}},}\ }\href {\doibase 10.1103/PhysRevLett.119.170501} {\bibfield
  {journal} {\bibinfo  {journal} {Physical Review Letters}\ }\textbf {\bibinfo
  {volume} {119}},\ \bibinfo {pages} {170501} (\bibinfo {year}
  {2017})}\BibitemShut {NoStop}%
\bibitem [{\citenamefont {Zhong}\ \emph {et~al.}(2021)\citenamefont {Zhong},
  \citenamefont {Deng}, \citenamefont {Qin}, \citenamefont {Wang},
  \citenamefont {Chen}, \citenamefont {Peng}, \citenamefont {Luo},
  \citenamefont {Wu}, \citenamefont {Gong}, \citenamefont {Su}, \citenamefont
  {Hu}, \citenamefont {Hu}, \citenamefont {Yang}, \citenamefont {Zhang},
  \citenamefont {Li}, \citenamefont {Li}, \citenamefont {Jiang}, \citenamefont
  {Gan}, \citenamefont {Yang}, \citenamefont {You}, \citenamefont {Wang},
  \citenamefont {Li}, \citenamefont {Liu}, \citenamefont {Renema},
  \citenamefont {Lu},\ and\ \citenamefont {Pan}}]{zhong2021}%
  \BibitemOpen
  \bibfield  {author} {\bibinfo {author} {\bibfnamefont {Han-Sen}\ \bibnamefont
  {Zhong}}, \bibinfo {author} {\bibfnamefont {Yu-Hao}\ \bibnamefont {Deng}},
  \bibinfo {author} {\bibfnamefont {Jian}\ \bibnamefont {Qin}}, \bibinfo
  {author} {\bibfnamefont {Hui}\ \bibnamefont {Wang}}, \bibinfo {author}
  {\bibfnamefont {Ming-Cheng}\ \bibnamefont {Chen}}, \bibinfo {author}
  {\bibfnamefont {Li-Chao}\ \bibnamefont {Peng}}, \bibinfo {author}
  {\bibfnamefont {Yi-Han}\ \bibnamefont {Luo}}, \bibinfo {author}
  {\bibfnamefont {Dian}\ \bibnamefont {Wu}}, \bibinfo {author} {\bibfnamefont
  {Si-Qiu}\ \bibnamefont {Gong}}, \bibinfo {author} {\bibfnamefont {Hao}\
  \bibnamefont {Su}}, \bibinfo {author} {\bibfnamefont {Yi}~\bibnamefont {Hu}},
  \bibinfo {author} {\bibfnamefont {Peng}\ \bibnamefont {Hu}}, \bibinfo
  {author} {\bibfnamefont {Xiao-Yan}\ \bibnamefont {Yang}}, \bibinfo {author}
  {\bibfnamefont {Wei-Jun}\ \bibnamefont {Zhang}}, \bibinfo {author}
  {\bibfnamefont {Hao}\ \bibnamefont {Li}}, \bibinfo {author} {\bibfnamefont
  {Yuxuan}\ \bibnamefont {Li}}, \bibinfo {author} {\bibfnamefont {Xiao}\
  \bibnamefont {Jiang}}, \bibinfo {author} {\bibfnamefont {Lin}\ \bibnamefont
  {Gan}}, \bibinfo {author} {\bibfnamefont {Guangwen}\ \bibnamefont {Yang}},
  \bibinfo {author} {\bibfnamefont {Lixing}\ \bibnamefont {You}}, \bibinfo
  {author} {\bibfnamefont {Zhen}\ \bibnamefont {Wang}}, \bibinfo {author}
  {\bibfnamefont {Li}~\bibnamefont {Li}}, \bibinfo {author} {\bibfnamefont
  {Nai-Le}\ \bibnamefont {Liu}}, \bibinfo {author} {\bibfnamefont {Jelmer~J.}\
  \bibnamefont {Renema}}, \bibinfo {author} {\bibfnamefont {Chao-Yang}\
  \bibnamefont {Lu}}, \ and\ \bibinfo {author} {\bibfnamefont {Jian-Wei}\
  \bibnamefont {Pan}},\ }\bibfield  {title} {\enquote {\bibinfo {title}
  {Phase-programmable gaussian boson sampling using stimulated squeezed
  light},}\ }\href {\doibase 10.1103/PhysRevLett.127.180502} {\bibfield
  {journal} {\bibinfo  {journal} {Phys. Rev. Lett.}\ }\textbf {\bibinfo
  {volume} {127}},\ \bibinfo {pages} {180502} (\bibinfo {year}
  {2021})}\BibitemShut {NoStop}%
\bibitem [{\citenamefont {Madsen}\ \emph {et~al.}(2022)\citenamefont {Madsen},
  \citenamefont {Laudenbach}, \citenamefont {Askarani}, \citenamefont
  {Rortais}, \citenamefont {Vincent}, \citenamefont {Bulmer}, \citenamefont
  {Miatto}, \citenamefont {Neuhaus}, \citenamefont {Helt}, \citenamefont
  {Collins}, \citenamefont {Lita}, \citenamefont {Gerrits}, \citenamefont
  {Nam}, \citenamefont {Vaidya}, \citenamefont {Menotti}, \citenamefont
  {Dhand}, \citenamefont {Vernon}, \citenamefont {Quesada},\ and\ \citenamefont
  {Lavoie}}]{madsen2022}%
  \BibitemOpen
  \bibfield  {author} {\bibinfo {author} {\bibfnamefont {Lars~S.}\ \bibnamefont
  {Madsen}}, \bibinfo {author} {\bibfnamefont {Fabian}\ \bibnamefont
  {Laudenbach}}, \bibinfo {author} {\bibfnamefont {Mohsen~Falamarzi.}\
  \bibnamefont {Askarani}}, \bibinfo {author} {\bibfnamefont {Fabien}\
  \bibnamefont {Rortais}}, \bibinfo {author} {\bibfnamefont {Trevor}\
  \bibnamefont {Vincent}}, \bibinfo {author} {\bibfnamefont {Jacob F.~F.}\
  \bibnamefont {Bulmer}}, \bibinfo {author} {\bibfnamefont {Filippo~M.}\
  \bibnamefont {Miatto}}, \bibinfo {author} {\bibfnamefont {Leonhard}\
  \bibnamefont {Neuhaus}}, \bibinfo {author} {\bibfnamefont {Lukas~G.}\
  \bibnamefont {Helt}}, \bibinfo {author} {\bibfnamefont {Matthew~J.}\
  \bibnamefont {Collins}}, \bibinfo {author} {\bibfnamefont {Adriana~E.}\
  \bibnamefont {Lita}}, \bibinfo {author} {\bibfnamefont {Thomas}\ \bibnamefont
  {Gerrits}}, \bibinfo {author} {\bibfnamefont {Sae~Woo}\ \bibnamefont {Nam}},
  \bibinfo {author} {\bibfnamefont {Varun~D.}\ \bibnamefont {Vaidya}}, \bibinfo
  {author} {\bibfnamefont {Matteo}\ \bibnamefont {Menotti}}, \bibinfo {author}
  {\bibfnamefont {Ish}\ \bibnamefont {Dhand}}, \bibinfo {author} {\bibfnamefont
  {Zachary}\ \bibnamefont {Vernon}}, \bibinfo {author} {\bibfnamefont
  {Nicol{\'a}s}\ \bibnamefont {Quesada}}, \ and\ \bibinfo {author}
  {\bibfnamefont {Jonathan}\ \bibnamefont {Lavoie}},\ }\bibfield  {title}
  {\enquote {\bibinfo {title} {Quantum computational advantage with a
  programmable photonic processor},}\ }\href {\doibase
  10.1038/s41586-022-04725-x} {\bibfield  {journal} {\bibinfo  {journal}
  {Nature}\ }\textbf {\bibinfo {volume} {606}},\ \bibinfo {pages} {75--81}
  (\bibinfo {year} {2022})}\BibitemShut {NoStop}%
\bibitem [{\citenamefont {Huh}\ \emph {et~al.}(2015)\citenamefont {Huh},
  \citenamefont {Guerreschi}, \citenamefont {Peropadre}, \citenamefont
  {McClean},\ and\ \citenamefont {{Aspuru-Guzik}}}]{huh2015}%
  \BibitemOpen
  \bibfield  {author} {\bibinfo {author} {\bibfnamefont {Joonsuk}\ \bibnamefont
  {Huh}}, \bibinfo {author} {\bibfnamefont {Gian~Giacomo}\ \bibnamefont
  {Guerreschi}}, \bibinfo {author} {\bibfnamefont {Borja}\ \bibnamefont
  {Peropadre}}, \bibinfo {author} {\bibfnamefont {Jarrod~R.}\ \bibnamefont
  {McClean}}, \ and\ \bibinfo {author} {\bibfnamefont {Al{\'a}n}\ \bibnamefont
  {{Aspuru-Guzik}}},\ }\bibfield  {title} {\enquote {\bibinfo {title} {Boson
  sampling for molecular vibronic spectra},}\ }\href {\doibase
  10.1038/nphoton.2015.153} {\bibfield  {journal} {\bibinfo  {journal} {Nature
  Photonics}\ }\textbf {\bibinfo {volume} {9}},\ \bibinfo {pages} {615--620}
  (\bibinfo {year} {2015})}\BibitemShut {NoStop}%
\bibitem [{\citenamefont {Arrazola}\ and\ \citenamefont
  {Bromley}(2018)}]{arrazola2018}%
  \BibitemOpen
  \bibfield  {author} {\bibinfo {author} {\bibfnamefont {Juan~Miguel}\
  \bibnamefont {Arrazola}}\ and\ \bibinfo {author} {\bibfnamefont {Thomas~R.}\
  \bibnamefont {Bromley}},\ }\bibfield  {title} {\enquote {\bibinfo {title}
  {Using {{Gaussian Boson Sampling}} to {{Find Dense Subgraphs}}},}\ }\href
  {\doibase 10.1103/PhysRevLett.121.030503} {\bibfield  {journal} {\bibinfo
  {journal} {Physical Review Letters}\ }\textbf {\bibinfo {volume} {121}},\
  \bibinfo {pages} {030503} (\bibinfo {year} {2018})}\BibitemShut {NoStop}%
\bibitem [{\citenamefont {Marek}\ \emph {et~al.}(2018)\citenamefont {Marek},
  \citenamefont {Filip}, \citenamefont {Ogawa}, \citenamefont {Sakaguchi},
  \citenamefont {Takeda}, \citenamefont {Yoshikawa},\ and\ \citenamefont
  {Furusawa}}]{marek2018a}%
  \BibitemOpen
  \bibfield  {author} {\bibinfo {author} {\bibfnamefont {Petr}\ \bibnamefont
  {Marek}}, \bibinfo {author} {\bibfnamefont {Radim}\ \bibnamefont {Filip}},
  \bibinfo {author} {\bibfnamefont {Hisashi}\ \bibnamefont {Ogawa}}, \bibinfo
  {author} {\bibfnamefont {Atsushi}\ \bibnamefont {Sakaguchi}}, \bibinfo
  {author} {\bibfnamefont {Shuntaro}\ \bibnamefont {Takeda}}, \bibinfo {author}
  {\bibfnamefont {Jun-ichi}\ \bibnamefont {Yoshikawa}}, \ and\ \bibinfo
  {author} {\bibfnamefont {Akira}\ \bibnamefont {Furusawa}},\ }\bibfield
  {title} {\enquote {\bibinfo {title} {General implementation of arbitrary
  nonlinear quadrature phase gates},}\ }\href {\doibase
  10.1103/PhysRevA.97.022329} {\bibfield  {journal} {\bibinfo  {journal}
  {Physical Review A}\ }\textbf {\bibinfo {volume} {97}},\ \bibinfo {pages}
  {022329} (\bibinfo {year} {2018})}\BibitemShut {NoStop}%
\bibitem [{\citenamefont {Zhou}\ \emph {et~al.}(2020)\citenamefont {Zhou},
  \citenamefont {Wang}, \citenamefont {Choi}, \citenamefont {Pichler},\ and\
  \citenamefont {Lukin}}]{zhou2020}%
  \BibitemOpen
  \bibfield  {author} {\bibinfo {author} {\bibfnamefont {Leo}\ \bibnamefont
  {Zhou}}, \bibinfo {author} {\bibfnamefont {Sheng-Tao}\ \bibnamefont {Wang}},
  \bibinfo {author} {\bibfnamefont {Soonwon}\ \bibnamefont {Choi}}, \bibinfo
  {author} {\bibfnamefont {Hannes}\ \bibnamefont {Pichler}}, \ and\ \bibinfo
  {author} {\bibfnamefont {Mikhail~D.}\ \bibnamefont {Lukin}},\ }\bibfield
  {title} {\enquote {\bibinfo {title} {Quantum {{Approximate Optimization
  Algorithm}}: {{Performance}}, {{Mechanism}}, and {{Implementation}} on
  {{Near-Term Devices}}},}\ }\href {\doibase 10.1103/PhysRevX.10.021067}
  {\bibfield  {journal} {\bibinfo  {journal} {Physical Review X}\ }\textbf
  {\bibinfo {volume} {10}},\ \bibinfo {pages} {021067} (\bibinfo {year}
  {2020})}\BibitemShut {NoStop}%
\bibitem [{\citenamefont {Willsch}\ \emph {et~al.}(2020)\citenamefont
  {Willsch}, \citenamefont {Willsch}, \citenamefont {Jin}, \citenamefont
  {De~Raedt},\ and\ \citenamefont {Michielsen}}]{willsch2020}%
  \BibitemOpen
  \bibfield  {author} {\bibinfo {author} {\bibfnamefont {Madita}\ \bibnamefont
  {Willsch}}, \bibinfo {author} {\bibfnamefont {Dennis}\ \bibnamefont
  {Willsch}}, \bibinfo {author} {\bibfnamefont {Fengping}\ \bibnamefont {Jin}},
  \bibinfo {author} {\bibfnamefont {Hans}\ \bibnamefont {De~Raedt}}, \ and\
  \bibinfo {author} {\bibfnamefont {Kristel}\ \bibnamefont {Michielsen}},\
  }\bibfield  {title} {\enquote {\bibinfo {title} {Benchmarking the quantum
  approximate optimization algorithm},}\ }\href {\doibase
  10.1007/s11128-020-02692-8} {\bibfield  {journal} {\bibinfo  {journal}
  {Quantum Information Processing}\ }\textbf {\bibinfo {volume} {19}},\
  \bibinfo {pages} {197} (\bibinfo {year} {2020})}\BibitemShut {NoStop}%
\bibitem [{\citenamefont {Cornu{\'e}jols}\ \emph {et~al.}(2006)\citenamefont
  {Cornu{\'e}jols}, \citenamefont {Pe{\~n}a},\ and\ \citenamefont
  {T{\"u}t{\"u}nc{\"u}}}]{cornuejols2006}%
  \BibitemOpen
  \bibfield  {author} {\bibinfo {author} {\bibfnamefont {G{\'e}rard}\
  \bibnamefont {Cornu{\'e}jols}}, \bibinfo {author} {\bibfnamefont {Javier}\
  \bibnamefont {Pe{\~n}a}}, \ and\ \bibinfo {author} {\bibfnamefont {Reha}\
  \bibnamefont {T{\"u}t{\"u}nc{\"u}}},\ }\href@noop {} {\emph {\bibinfo {title}
  {Optimization Methods in Finance}}}\ (\bibinfo  {publisher} {{Cambridge
  University Press}},\ \bibinfo {year} {2006})\BibitemShut {NoStop}%
\bibitem [{\citenamefont {Aggarwal}(2020)}]{aggarwal2020}%
  \BibitemOpen
  \bibfield  {author} {\bibinfo {author} {\bibfnamefont {Charu~C.}\
  \bibnamefont {Aggarwal}},\ }\href {\doibase 10.1007/978-3-030-40344-7} {\emph
  {\bibinfo {title} {Linear {{Algebra}} and {{Optimization}} for {{Machine
  Learning}}}}}\ (\bibinfo  {publisher} {{Springer International Publishing}},\
  \bibinfo {address} {{Cham}},\ \bibinfo {year} {2020})\BibitemShut {NoStop}%
\bibitem [{\citenamefont {Rao}(1996)}]{rao1996}%
  \BibitemOpen
  \bibfield  {author} {\bibinfo {author} {\bibfnamefont {Singiresu~S.}\
  \bibnamefont {Rao}},\ }\href@noop {} {\emph {\bibinfo {title} {Engineering
  Optimization : Theory and Practice}}}\ (\bibinfo  {publisher} {{John Wiley \&
  Sons}},\ \bibinfo {year} {1996})\BibitemShut {NoStop}%
\bibitem [{\citenamefont {Pati}\ \emph {et~al.}(2000)\citenamefont {Pati},
  \citenamefont {Braunstein},\ and\ \citenamefont {Lloyd}}]{pati2000}%
  \BibitemOpen
  \bibfield  {author} {\bibinfo {author} {\bibfnamefont {Arun~K.}\ \bibnamefont
  {Pati}}, \bibinfo {author} {\bibfnamefont {Samuel~L.}\ \bibnamefont
  {Braunstein}}, \ and\ \bibinfo {author} {\bibfnamefont {Seth}\ \bibnamefont
  {Lloyd}},\ }\bibfield  {title} {\enquote {\bibinfo {title} {Quantum searching
  with continuous variables},}\ }\href@noop {} {\bibfield  {journal} {\bibinfo
  {journal} {arXiv:quant-ph/0002082}\ } (\bibinfo {year} {2000})},\ \Eprint
  {http://arxiv.org/abs/quant-ph/0002082} {arXiv:quant-ph/0002082} \BibitemShut
  {NoStop}%
\bibitem [{\citenamefont {Fukui}\ and\ \citenamefont
  {Takeda}(2022)}]{fukui2022}%
  \BibitemOpen
  \bibfield  {author} {\bibinfo {author} {\bibfnamefont {Kosuke}\ \bibnamefont
  {Fukui}}\ and\ \bibinfo {author} {\bibfnamefont {Shuntaro}\ \bibnamefont
  {Takeda}},\ }\bibfield  {title} {\enquote {\bibinfo {title} {Building a
  large-scale quantum computer with continuous-variable optical
  technologies},}\ }\href {\doibase 10.1088/1361-6455/ac489c} {\bibfield
  {journal} {\bibinfo  {journal} {Journal of Physics B}\ }\textbf {\bibinfo
  {volume} {55}},\ \bibinfo {pages} {012001} (\bibinfo {year}
  {2022})}\BibitemShut {NoStop}%
\bibitem [{\citenamefont {Asavanant}\ \emph {et~al.}(2021)\citenamefont
  {Asavanant}, \citenamefont {Charoensombutamon}, \citenamefont {Yokoyama},
  \citenamefont {Ebihara}, \citenamefont {Nakamura}, \citenamefont {Alexander},
  \citenamefont {Endo}, \citenamefont {Yoshikawa}, \citenamefont {Menicucci},
  \citenamefont {Yonezawa},\ and\ \citenamefont {Furusawa}}]{asavanant2021a}%
  \BibitemOpen
  \bibfield  {author} {\bibinfo {author} {\bibfnamefont {Warit}\ \bibnamefont
  {Asavanant}}, \bibinfo {author} {\bibfnamefont {Baramee}\ \bibnamefont
  {Charoensombutamon}}, \bibinfo {author} {\bibfnamefont {Shota}\ \bibnamefont
  {Yokoyama}}, \bibinfo {author} {\bibfnamefont {Takeru}\ \bibnamefont
  {Ebihara}}, \bibinfo {author} {\bibfnamefont {Tomohiro}\ \bibnamefont
  {Nakamura}}, \bibinfo {author} {\bibfnamefont {Rafael~N.}\ \bibnamefont
  {Alexander}}, \bibinfo {author} {\bibfnamefont {Mamoru}\ \bibnamefont
  {Endo}}, \bibinfo {author} {\bibfnamefont {Jun-ichi}\ \bibnamefont
  {Yoshikawa}}, \bibinfo {author} {\bibfnamefont {Nicolas~C.}\ \bibnamefont
  {Menicucci}}, \bibinfo {author} {\bibfnamefont {Hidehiro}\ \bibnamefont
  {Yonezawa}}, \ and\ \bibinfo {author} {\bibfnamefont {Akira}\ \bibnamefont
  {Furusawa}},\ }\bibfield  {title} {\enquote {\bibinfo {title}
  {Time-{{Domain-Multiplexed Measurement-Based Quantum Operations}} with
  25-{{MHz Clock Frequency}}},}\ }\href {\doibase
  10.1103/PhysRevApplied.16.034005} {\bibfield  {journal} {\bibinfo  {journal}
  {Physical Review Applied}\ }\textbf {\bibinfo {volume} {16}},\ \bibinfo
  {pages} {034005} (\bibinfo {year} {2021})},\ \Eprint
  {http://arxiv.org/abs/2006.11537} {arXiv:2006.11537} \BibitemShut {NoStop}%
\bibitem [{\citenamefont {Larsen}\ \emph {et~al.}(2021)\citenamefont {Larsen},
  \citenamefont {Guo}, \citenamefont {Breum}, \citenamefont
  {{Neergaard-Nielsen}},\ and\ \citenamefont {Andersen}}]{larsen2021a}%
  \BibitemOpen
  \bibfield  {author} {\bibinfo {author} {\bibfnamefont {Mikkel~V.}\
  \bibnamefont {Larsen}}, \bibinfo {author} {\bibfnamefont {Xueshi}\
  \bibnamefont {Guo}}, \bibinfo {author} {\bibfnamefont {Casper~R.}\
  \bibnamefont {Breum}}, \bibinfo {author} {\bibfnamefont {Jonas~S.}\
  \bibnamefont {{Neergaard-Nielsen}}}, \ and\ \bibinfo {author} {\bibfnamefont
  {Ulrik~L.}\ \bibnamefont {Andersen}},\ }\bibfield  {title} {\enquote
  {\bibinfo {title} {Deterministic multi-mode gates on a scalable photonic
  quantum computing platform},}\ }\href {\doibase 10.1038/s41567-021-01296-y}
  {\bibfield  {journal} {\bibinfo  {journal} {Nature Physics}\ }\textbf
  {\bibinfo {volume} {17}},\ \bibinfo {pages} {1018--1023} (\bibinfo {year}
  {2021})},\ \Eprint {http://arxiv.org/abs/2010.14422} {arXiv:2010.14422}
  \BibitemShut {NoStop}%
\bibitem [{\citenamefont {Enomoto}\ \emph {et~al.}(2021)\citenamefont
  {Enomoto}, \citenamefont {Yonezu}, \citenamefont {Mitsuhashi}, \citenamefont
  {Takase},\ and\ \citenamefont {Takeda}}]{enomoto2021}%
  \BibitemOpen
  \bibfield  {author} {\bibinfo {author} {\bibfnamefont {Yutaro}\ \bibnamefont
  {Enomoto}}, \bibinfo {author} {\bibfnamefont {Kazuma}\ \bibnamefont
  {Yonezu}}, \bibinfo {author} {\bibfnamefont {Yosuke}\ \bibnamefont
  {Mitsuhashi}}, \bibinfo {author} {\bibfnamefont {Kan}\ \bibnamefont
  {Takase}}, \ and\ \bibinfo {author} {\bibfnamefont {Shuntaro}\ \bibnamefont
  {Takeda}},\ }\bibfield  {title} {\enquote {\bibinfo {title} {Programmable and
  sequential {{Gaussian}} gates in a loop-based single-mode photonic quantum
  processor},}\ }\href {\doibase 10.1126/sciadv.abj6624} {\bibfield  {journal}
  {\bibinfo  {journal} {Science Advances}\ }\textbf {\bibinfo {volume} {7}},\
  \bibinfo {pages} {eabj6624} (\bibinfo {year} {2021})}\BibitemShut {NoStop}%
\bibitem [{\citenamefont {Yoshikawa}\ \emph {et~al.}(2016)\citenamefont
  {Yoshikawa}, \citenamefont {Yokoyama}, \citenamefont {Kaji}, \citenamefont
  {Sornphiphatphong}, \citenamefont {Shiozawa}, \citenamefont {Makino},\ and\
  \citenamefont {Furusawa}}]{yoshikawa2016}%
  \BibitemOpen
  \bibfield  {author} {\bibinfo {author} {\bibfnamefont {Junichi}\ \bibnamefont
  {Yoshikawa}}, \bibinfo {author} {\bibfnamefont {Shota}\ \bibnamefont
  {Yokoyama}}, \bibinfo {author} {\bibfnamefont {Toshiyuki}\ \bibnamefont
  {Kaji}}, \bibinfo {author} {\bibfnamefont {Chanond}\ \bibnamefont
  {Sornphiphatphong}}, \bibinfo {author} {\bibfnamefont {Yu}~\bibnamefont
  {Shiozawa}}, \bibinfo {author} {\bibfnamefont {Kenzo}\ \bibnamefont
  {Makino}}, \ and\ \bibinfo {author} {\bibfnamefont {Akira}\ \bibnamefont
  {Furusawa}},\ }\bibfield  {title} {\enquote {\bibinfo {title} {Invited
  {{Article}}: {{Generation}} of one-million-mode continuous-variable cluster
  state by unlimited time-domain multiplexing},}\ }\href {\doibase
  10.1063/1.4962732} {\bibfield  {journal} {\bibinfo  {journal} {APL
  Photonics}\ }\textbf {\bibinfo {volume} {1}},\ \bibinfo {pages} {060801}
  (\bibinfo {year} {2016})}\BibitemShut {NoStop}%
\bibitem [{\citenamefont {Bartlett}\ \emph {et~al.}(2002)\citenamefont
  {Bartlett}, \citenamefont {Sanders}, \citenamefont {Braunstein},\ and\
  \citenamefont {Nemoto}}]{ths:ClassicalSimulation}%
  \BibitemOpen
  \bibfield  {author} {\bibinfo {author} {\bibfnamefont {Stephen~D.}\
  \bibnamefont {Bartlett}}, \bibinfo {author} {\bibfnamefont {Barry~C.}\
  \bibnamefont {Sanders}}, \bibinfo {author} {\bibfnamefont {Samuel~L.}\
  \bibnamefont {Braunstein}}, \ and\ \bibinfo {author} {\bibfnamefont {Kae}\
  \bibnamefont {Nemoto}},\ }\bibfield  {title} {\enquote {\bibinfo {title}
  {Efficient classical simulation of continuous variable quantum information
  processes},}\ }\href {\doibase 10.1103/PhysRevLett.88.097904} {\bibfield
  {journal} {\bibinfo  {journal} {Phys. Rev. Lett.}\ }\textbf {\bibinfo
  {volume} {88}},\ \bibinfo {pages} {097904} (\bibinfo {year}
  {2002})}\BibitemShut {NoStop}%
\bibitem [{\citenamefont {Arai}\ \emph {et~al.}(2023)\citenamefont {Arai},
  \citenamefont {Oshiyama},\ and\ \citenamefont {Nishimori}}]{ths:DVforCV}%
  \BibitemOpen
  \bibfield  {author} {\bibinfo {author} {\bibfnamefont {Shunta}\ \bibnamefont
  {Arai}}, \bibinfo {author} {\bibfnamefont {Hiroki}\ \bibnamefont {Oshiyama}},
  \ and\ \bibinfo {author} {\bibfnamefont {Hidetoshi}\ \bibnamefont
  {Nishimori}},\ }\href@noop {} {\enquote {\bibinfo {title} {Quantum annealing
  for continuous-variable optimization: How is it effective?}}\ } (\bibinfo
  {year} {2023}),\ \Eprint {http://arxiv.org/abs/2305.06631} {arXiv:2305.06631
  [quant-ph]} \BibitemShut {NoStop}%
\bibitem [{\citenamefont {Miyata}\ \emph {et~al.}(2014)\citenamefont {Miyata},
  \citenamefont {Ogawa}, \citenamefont {Marek}, \citenamefont {Filip},
  \citenamefont {Yonezawa}, \citenamefont {Yoshikawa},\ and\ \citenamefont
  {Furusawa}}]{miyata2014}%
  \BibitemOpen
  \bibfield  {author} {\bibinfo {author} {\bibfnamefont {Kazunori}\
  \bibnamefont {Miyata}}, \bibinfo {author} {\bibfnamefont {Hisashi}\
  \bibnamefont {Ogawa}}, \bibinfo {author} {\bibfnamefont {Petr}\ \bibnamefont
  {Marek}}, \bibinfo {author} {\bibfnamefont {Radim}\ \bibnamefont {Filip}},
  \bibinfo {author} {\bibfnamefont {Hidehiro}\ \bibnamefont {Yonezawa}},
  \bibinfo {author} {\bibfnamefont {Jun-ichi}\ \bibnamefont {Yoshikawa}}, \
  and\ \bibinfo {author} {\bibfnamefont {Akira}\ \bibnamefont {Furusawa}},\
  }\bibfield  {title} {\enquote {\bibinfo {title} {Experimental realization of
  a dynamic squeezing gate},}\ }\href {\doibase 10.1103/PhysRevA.90.060302}
  {\bibfield  {journal} {\bibinfo  {journal} {Physical Review A}\ }\textbf
  {\bibinfo {volume} {90}},\ \bibinfo {pages} {060302} (\bibinfo {year}
  {2014})}\BibitemShut {NoStop}%
\bibitem [{\citenamefont {Nogueira}(2014--)}]{package}%
  \BibitemOpen
  \bibfield  {author} {\bibinfo {author} {\bibfnamefont {Fernando}\
  \bibnamefont {Nogueira}},\ }\href
  {https://github.com/fmfn/BayesianOptimization} {\enquote {\bibinfo {title}
  {{Bayesian Optimization}: Open source constrained global optimization tool
  for {Python}},}\ } (\bibinfo {year} {2014--})\BibitemShut {NoStop}%
\end{thebibliography}

\end{document}